%% file: main.tex
\documentclass[manuscript,nonacm]{acmart}

\usepackage{array}
\usepackage{acmart-taps} 

\usepackage{booktabs}
\usepackage{caption}
\usepackage{comment}

\usepackage{cleveref}
\usepackage{color}
\usepackage{enumitem}
\usepackage{float}
\usepackage{fancybox}
\usepackage{graphicx} 
\usepackage{xcolor}
\usepackage{booktabs}
\usepackage{tabularx}
\usepackage{longtable}
\usepackage{multirow}
\usepackage{listings}

\usepackage{xspace}

\lstdefinestyle{classificationprompt}{
    basicstyle=\ttfamily\small,
    columns=fullflexible,
    keepspaces=true,
    breaklines=true,
    breakatwhitespace=false,
    frame=single,
    framesep=6pt,
    rulecolor=\color{black!35},
    backgroundcolor=\color{gray!4},
    showstringspaces=false,
    captionpos=t,
    literate={—}{{\textemdash}}1
}

\crefname{section}{Section}{Sections}
\crefname{subsection}{Section}{Sections}
\crefname{subsubsection}{Section}{Sections}

\AtBeginDocument{}

\begin{document}

\title[What should AI forget?]{What should AI forget? A design and taxonomy for negative space governance in AI memory}

\author{Eryue Xu}
\affiliation{
  \institution{University of Illinois Urbana-Champaign}
  \city{Champaign}
  \state{Illinois}
  \country{USA}}
\email{eryuexu2@illinois.edu}

  \author{Ke Ding}
\affiliation{
  \institution{University of Illinois Urbana-Champaign}
  \city{Champaign}
  \state{Illinois}
  \country{USA}}
  \email{keding4@illinois.edu}

  \author{Ting-Yu Chang}
\affiliation{
  \institution{University of Illinois Urbana-Champaign}
  \city{Champaign}
  \state{Illinois}
  \country{USA}}
  \email{tc86@illinois.edu}  

  \author{Tianshi Li}
\affiliation{
  \institution{Northeastern University}
  \city{Boston}
  \state{Massachusetts}
  \country{USA}}
  \email{tia.li@northeastern.edu} 

  \author{Masooda Bashir}
\affiliation{
  \institution{University of Illinois Urbana-Champaign}
  \city{Champaign}
  \state{Illinois}
  \country{USA}}
  \email{mnb@illinois.edu}

\renewcommand{\shortauthors}{Xu et al.}

\begin{abstract}

AI memory personalizes assistants, yet users want some disclosed information forgotten. We characterize the negative space of AI memory, where protection should take precedence over potential future personalization benefits. From 179 incidents reported by 146 users, we developed a taxonomy of 15 information categories users want excluded from memory, revealing sensitivity concerns beyond conventional taxonomies. We operationalize this taxonomy in FENCE, a policy-level memory gate that automatically applies user-selected category-based rules to subsequent conversations, preventing information from protected conversations from entering AI memory by default while allowing case-by-case overrides. In a within-subjects study (N=15), participants described FENCE as making memory governance more usable, interpretable, and trustworthy than conventional post-hoc management. Our findings demonstrate how taxonomy-based rules can potentially support user-defined memory boundaries. We call for collective efforts to establish policies governing what AI should not remember, rather than leaving protection to individual retrospective management.

\end{abstract}

\ccsdesc[300]{Human-centered computing~User studies}
\ccsdesc[500]{Security and privacy~Social aspects of security and privacy}
\ccsdesc[300]{Computing methodologies~Intelligent agents}

\keywords{AI memory, Conversational AI agents, Usable Privacy, Privacy Taxonomy, System Design, Memory Governance}

\begin{teaserfigure}
    \centering
    \includegraphics[width=\textwidth]{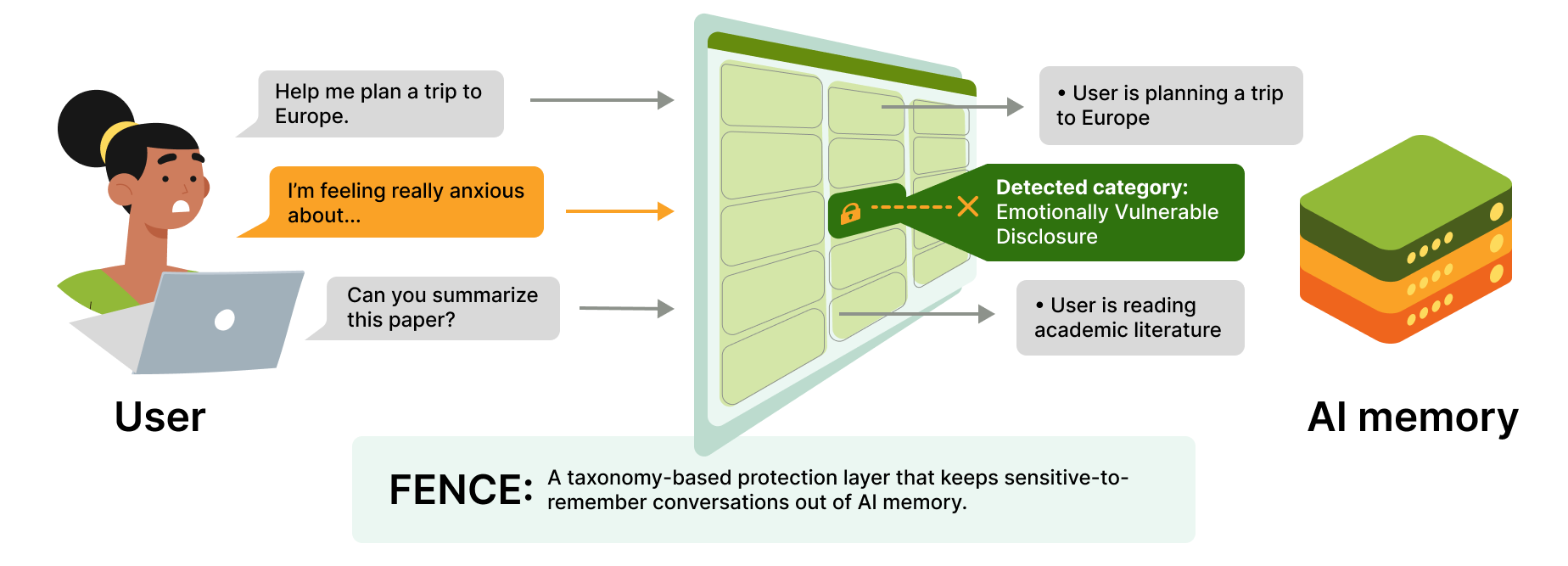}
    \caption{
        Conceptual overview of \textit{FENCE}. 
        \textit{FENCE} applies user-defined category-based policies as information transitions from conversation to persistent AI memory. 
        Disclosures covered by the policy are kept out of memory, while other information remains eligible for persistent memory.
    }
    \Description{
        Conceptual illustration of FENCE as a category-based policy layer between conversation and persistent AI memory. 
        A user's ordinary messages about planning a trip to Europe and reading academic literature pass through FENCE and remain eligible for memory. 
        An emotionally vulnerable disclosure is detected as belonging to a protected category and is prevented from entering persistent AI memory.
    }
    \label{fig:fence-concept}
\end{teaserfigure}

\maketitle

\newcommand{\fence}{\textsc{Fence}\xspace}

\input{sections/1_Introduction}
\input{sections/2_Related_Work}
\input{sections/3_Method}
\input{sections/4_Taxonomy}

\input{sections/5_Design}

\input{sections/5_Model_benchmark}
\input{sections/6_User_evaluation}
\input{sections/7_Discussion}

\bibliographystyle{ACM-Reference-Format}
\bibliography{bibliography}

\newpage
\input{sections/appendix}

\end{document}

%% file: sections/1_Introduction.tex
\section{Introduction}

Ever since AI began to speak like a human, we have hoped it might also know us as a person would: remembering who we are, understanding what we need, and personalizing its help over time.
AI memory sits at the center of this vision: persistent knowledge about the user provides the basis for more personalized and capable assistance. 
Researchers have therefore sought both to retain more useful context over time, from longer context windows to explicit cross-session memory mechanisms~\cite{Packer2023MemGPT,Zhong2024MemoryBank,Zhang2025MemorySurvey}, and to reuse retained information to better support future interactions at both technical~\cite{Park2023GenerativeAgents,Chhikara2025Mem0,Xu2025AMem} and usability~\cite{Huang2023MemorySandbox,Yen2024Memolet,Vaithilingam2025SemanticCommit} levels.
Across these efforts, user-disclosed information is largely treated as a pool of candidate memories: the main questions are what information is useful to retain and how that information should be used later.
We advocate for a complementary line of inquiry that has received less attention: \textit{what should AI not remember in the first place?} More specifically, what information do users want a general-purpose AI assistant to keep out of persistent memory? In this paper, we call this broader problem \textbf{\textit{memory governance}}.

Memory governance extends beyond conventional data-storage privacy concerns in several ways; it is not simply a matter of filtering out personally identifiable information. First, conversational AI makes free-form natural language its primary interaction channel, allowing users to disclose more diverse information~\cite{chatterji2025people, Tamkin2024Clio, CostaGomes2025Copilot, Phang2025Affective} and encouraging sensitive disclosure that users might otherwise be more cautious about sharing~\cite{Lucas2014Disclosure, Ho2018ChatbotDisclosure, zhang2024fair}. Second, AI memory mechanisms can extract information from conversations and continuously update representations of the user~\cite{Chhikara2025Mem0,Xu2025AMem} while LLMs can infer personal attributes that users never explicitly disclosed~\cite{Staab2024BeyondMemorization}.
Third, the eventual use and risk of a disclosure are therefore difficult to anticipate at the time of sharing. Persistent-memory attacks have been demonstrated in research and deployment~\cite{Wang2025MemoryPrivacy,Herrador2026SpAIware}, while tool use and connected accounts create new paths into and out of memory~\cite{Gao2026IsolatedExposed}. This leaves both users and researchers facing substantial uncertainty about how persistent memory may be used and misused~\cite{Malki2026UsableMemory,Patlan2025ContextManipulation,Sun2025MSA}.

We cannot simply ask users not to share personal information with AI.
General-purpose AI is already largely used across both work and everyday life for information seeking, practical guidance, and decision support~\cite{Tamkin2024Clio, CostaGomes2025Copilot, Phang2025Affective}. In many of these uses, information about the user's situation is precisely what allows the system to provide a useful response. Privacy-preserving work has therefore increasingly framed the goal not as avoiding disclosure, but as minimizing disclosure while preserving what is necessary for the user's task~\cite{Ngong2025ContextualPrivacy, Zhou2025Rescriber, zhou2026operationalizing}. We learned from this body of research that while users may sometimes disclose information without recognizing its sensitivity, they may also knowingly disclose sensitive context because they need the AI to process it.
This raises a different design requirement: can information be available to the AI for the task at hand while still being restricted from becoming part of its persistent memory?

Existing AI memory controls only partially support this distinction.

At the account level, users can disable memory or cross-chat reference altogether~\cite{OpenAI2026MemoryFAQ,Google2026GeminiMemory,Anthropic2026ClaudeMemory}, but this broad exclusion sacrifices personalization even for information users want the system to remember. At the item level, users can return to memory settings, chat history, or activity pages to inspect and remove information that has already been retained~\cite{OpenAI2026MemoryFAQ,Google2026GeminiMemory,Anthropic2026ClaudeMemory}. This offers finer control, but requires users to discover what was remembered and manually identify what should be corrected or forgotten. The closest existing option is session-level protection through Temporary Chat in ChatGPT and Gemini or Incognito Chat in Claude~\cite{OpenAI2026TemporaryChat,Google2026TemporaryChat,Anthropic2026Incognito}. These modes selectively keep a conversation from contributing to future memory, but they must be chosen when starting the conversation. 
If users realize only after sharing that some information should not be remembered, temporary mode is no longer capable of protecting that disclosure.
 These limitations leave an important post-sharing gap. Effective protection requires acting before disclosure via data minimization, self-censorship, or temporary chat; once information has been shared, users struggle to find usable ways to govern memory.
 
In this paper, we explore the opportunity to establish a standing rule that governs AI memory over time, an approach we call \textbf{\textit{policy-level protection}}. Following Yang et al.'s framing of policy as principles, guidelines, and rules together with how they are enacted through systems~\cite{Yang2024FutureHCIPolicy}, \textit{policy-level} means that the unit of protection is a set of rules governing system behavior rather than an individual memory object. Such a set of rules specifies how a class of information should be handled and continues to apply whenever similar cases arise. In our case, the rules determine which conversations cannot contribute information to persistent memory, without requiring users to recognize the need for protection before sharing or manually delete the resulting memories afterward. The information that users disclose but want to exclude from persistent memory constitutes what we call the \textbf{\textit{negative space of AI memory}}. Making this protection actionable requires a vocabulary for identifying which kinds of information belong in this space and should be governed by the same rule. Therefore, we ask:

\begin{itemize}
\item \textbf{RQ1} What kinds of information have users shared but want a conversational AI assistant to forget?
\item \textbf{RQ2} How can we build a usable module that allows the AI system to intelligently gate this information out of its memory?
\end{itemize}

To answer RQ1, we designed a survey asking users what they had disclosed to an AI assistant but would prefer it to forget. Asking what users want the AI to \textit{forget} gives us a post-sharing view of the negative space of AI memory by identifying information that users disclosed for an immediate interaction but did not want retained afterward.

We collected 179 what-AI-to-forget incidents from 146 participants.
Through iterative coding, we developed a taxonomy of 15 categories of information that users want AI to forget.
Our taxonomy provides an empirical characterization of the \emph{negative space of AI memory}, grounded in users' actual experiences and desires. It extends beyond conventional sensitive-information categories~\cite{Milne2017InformationSensitivity} and those explicitly identified in platform memory policies~\cite{Anthropic2026ClaudeMemory} to include contextually sensitive disclosures, such as emotionally vulnerable experiences. These findings suggest that institutionally defined sensitivity boundaries do not fully capture what users consider inappropriate for an AI to remember.

To answer RQ2, we operationalize the taxonomy as a \textit{policy-level protection} layer in \textsc{Fence} (Filtered Entry of seNsitive Conversation to mEmory), a memory gate module that keeps conversations covered by a user's selected categories out of the AI's memory.
Once configured, \textsc{Fence} automatically detects disclosures in protected categories and prevents information from those conversations from entering persistent memory while preserving its use in the current interaction. These standing rules protect the user's selected negative space by default across subsequent conversations.
We ran a within-subjects study with 15 participants comparing \textsc{Fence} against conventional post-hoc memory management on the same tasks. Participants found our \textsc{Fence} system to be more usable and trustworthy and reported a substantially clearer understanding of what the system would remember and why. These benefits did not come simply from reducing effort; rather, participants valued shifting effort from repeatedly managing individual disclosures to setting a standing protection rule upfront.

Our work makes three contributions. First, we present an empirical taxonomy of 15 categories of information users want to keep out of AI memory, grounded in 179 real-life incidents. To our knowledge, this is the first systematic characterization of the negative space of AI memory from users' own accounts.

Second, we present \fence, a policy-level memory protection module that operationalizes the taxonomy as a standing rule and makes protection the default behavior for user-selected categories.
Third, we provide empirical evidence from a within-subjects study showing how policy-level protection can make memory governance feel more usable, trustworthy, and interpretable than post-hoc management.

Building on these findings, we argue that information usefulness in an immediate interaction does not by itself justify its persistent retention. We discuss how effective memory governance should protect this \textit{negative space} by default while making protection verifiable and allowing users to revise its scope. We further call for collective efforts to establish shared policies that govern what AI should not remember and how these protections should be enforced across systems.

%% file: sections/2_Related_Work.tex
\section{Related Work}

\subsection{User-facing AI Memory and Memory Management}

At the user-facing level, AI memory broadly refers to mechanisms through which an AI system retains information from previous interactions and makes it available to shape subsequent behavior.
As such mechanisms become standard in general-purpose AI assistants, HCI research has examined how users experience and manage them. 
Users often hold incomplete mental models of what AI systems remember, how memories are recalled, and how they influence system behavior \cite{Jones2025AgentMemory, Jones2025StorageInterpretation}. A growing set of tools and interfaces therefore supports user interaction with memory \cite{Kim2026MetaphorsMemory}. These tools follow two main paradigms. The first makes memory \textit{legible}, allowing users to view and manipulate what the system retains \cite{Huang2023MemorySandbox}; such visibility also shapes how users judge memory-enabled chatbots, as memory without visualization lowered users' assessments, whereas visualized memory improved them \cite{Yan2025RememberingTrust}. The second helps users \textit{curate memory for reuse}: Memolet lets users select information from prior conversations and specify how it should be reused \cite{Yen2024Memolet}, while Semantic Commit helps users update persistent information and resolve conflicts as their intentions change \cite{Vaithilingam2025SemanticCommit}. Across both paradigms, memory is primarily approached as a resource worth carrying forward: the design questions concern what to remember and reuse and how that information should be represented, organized, or updated.

Privacy research challenges this assumption, showing that retained information can become a liability rather than a resource. Users perceive AI memory as overly detailed and ``unforgetful,'' are often surprised by what it has retained, and grow uncomfortable when remembered or inferred information feels inaccurate, misrepresentative, or misaligned with how they expect it to be used \cite{Chen2026AINeverForgets, Zhang2025RAGMemoryPrivacy, Monteiro2026InferencesAcceptable}. Addressing these problems is left largely to users, who often misunderstand what deleting data actually accomplishes \cite{Malki2025Hoovered} and therefore call for greater transparency and more granular control over what is remembered and how it is used \cite{Zhang2025RAGMemoryPrivacy, Chen2026AINeverForgets}. Research has accordingly sought to understand and improve such control, ranging from studies characterizing how platforms let users access, edit, delete, and constrain memory \cite{Li2026PrivacyControl} to systems that proactively flag sensitive information in memory for users to edit or delete \cite{Zhang2024GhostOfThePast}. These efforts, however, primarily help users see and act on what has already been retained.

What remains open is a content-level question that precedes such control: \textit{what kinds of information should AI not remember in the first place?} We refer to the focus of prior work as the \textit{positive space} of memory---what is worth carrying forward and how to manage it, including when to delete it. Our work aims to explore the \textit{negative space}: information that should be excluded from AI memory by default. Without an account of this boundary, transparency relies on users to notice what should not have been retained, and control relies on them to repair it one memory at a time. We address this gap by empirically characterizing the negative space of AI memory and using the resulting taxonomy in a system that prevents such information from entering memory.

\subsection{Taxonomy on Sensitivity and Privacy}
Prior work has developed taxonomies of privacy at multiple levels. Some influential frameworks classify privacy problems according to the activities or information flows through which privacy can be compromised, such as collection, processing, dissemination, and inappropriate transmission across contexts \cite{Solove2006Taxonomy,Nissenbaum2004Privacy,Nissenbaum2019ContextualIntegrity}. While these frameworks are important for understanding privacy harms more broadly, our focus is on taxonomies that characterize the sensitivity of the \textit{information itself}.

One common approach is to define sensitivity through categories that warrant heightened protection. Privacy regulations operationalize this idea by enumerating information types that are treated differently from ordinary personal data. The GDPR, for example, identifies special categories including racial or ethnic origin, political and religious beliefs, biometric and health data, and information concerning sex life or sexual orientation, while the CCPA additionally highlights information such as precise geolocation, account credentials, private communications, and financial information \cite{GDPR2016,CCPA}. These classifications provide concrete protection boundaries, but academic work has repeatedly shown that sensitivity does not map neatly onto a fixed list of information types. Reviews of sensitive information emphasize that judgments vary with the potential consequences of disclosure and the circumstances in which information is used \cite{Thompson2005Sensitive,Ohm2015Sensitive}. Empirical taxonomies similarly organize information not only by semantic category but by the risks people associate with its disclosure. For instance, Milne et al.\ grouped 52 types of personal information according to perceived physical, psychological, monetary, and social risk, revealing substantial variation even among information commonly labeled as personal \cite{Milne2017InformationSensitivity}. Broader syntheses likewise find considerable disagreement across regulation, industry, and academic research over what should count as personal or sensitive information \cite{Saglam2022PersonalInformation}. Together, this work suggests a progression from treating sensitivity as a predefined property of certain data categories toward understanding it as a classification shaped by both information content and the harms that its disclosure may enable.

Conversational AI further stretches these boundaries~\cite{Zaman2026InferentialPrivacy, Mireshghallah2024KeepSecret, Tran2025PrivacyNorms, Li2025PrivaCIBench}. \citet{Mireshghallah2024TrustNoBot} show that real human--LLM conversations contain sensitive disclosures, such as sexual preferences or drug-use behaviors, that are poorly captured by conventional PII categories. AI memory introduces an additional complication as disclosed information may persist and become part of an evolving representation that shapes future interactions. This suggests that a taxonomy for AI memory should not be limited to conventionally defined sensitive information, but should reflect what users themselves want excluded from persistent memory.

%% file: sections/3_Method.tex
\section{Method}
\label{sec:method}

Our study consisted of two stages. We first conducted a survey to collect situations in which users wanted an AI to forget previously shared information and used these incidents to develop a taxonomy of sensitive-to-remember information. We then operationalized the taxonomy in \fence  and conducted a within-subjects study comparing \fence  with conventional post-hoc memory management. All participants were U.S.-based, English-speaking adults. The study protocol was approved by our institutional review board, and all participants provided informed consent.

\subsection{Survey to Taxonomy Development}

We conducted a qualitative survey to ground the taxonomy in users' experiences of wanting previously disclosed information forgotten rather than in predefined categories of sensitive information.

\subsubsection{Survey Design}

We adapted the Critical Incident Technique (CIT)~\citep{Flanagan1954CIT} to elicit specific experiences of wanting an AI to forget disclosed information rather than hypothetical judgments about what might be sensitive. CIT has similarly been used in privacy research to study disclosures people later regretted~\cite{Wang2011Regret}.

To support recall, participants first reported the AI platforms they used and reviewed their past uses through a checklist adapted from~\citet{chatterji2025people}. We then displayed their selections while asking them to recall a relevant incident. For each incident, participants described the interaction context, what information they disclosed, their expectations at the time, and when and why they later wanted the information forgotten. We explicitly asked participants not to reproduce the exact sensitive messages they had shared. Optional questions asked about strategies they had used to address the concern, their experience with those strategies, and desired alternatives. We disabled copy-and-paste in open-response fields to discourage AI-generated responses. The full questionnaire is provided in Appendix~\ref{app:formative-survey}.

\subsubsection{Participants}

We recruited participants through Prolific~\citep{PalanSchitter2018Prolific}. Eligibility required at least one month of conversational AI use and a prior experience of wanting an AI to forget information they had shared. Participants who did not meet these criteria were screened out before the incident questions and received \$0.18.
Participants were required to report one incident and could optionally report up to two additional incidents. After excluding eight responses that did not provide meaningful incident descriptions (e.g., ``N/A''), the final dataset contained 179 incidents from 146 participants. The survey took an average of six minutes. Participants received \$1 for completion and a single \$0.50 bonus for providing one or more additional incidents.

\subsubsection{Developing the Taxonomy}

We analyzed the incidents through iterative qualitative coding. Our primary unit of analysis was the \textit{information being protected}: what kind of information the participant did not want retained, rather than the threat or harm they anticipated.
Coding proceeded in two phases. First, three researchers independently open-coded the same 30 incidents and then met to compare and consolidate their codes. This produced an initial taxonomy of 14 sensitive-to-remember categories with working definitions. Second, two researchers independently applied this taxonomy to the remaining incidents. Incidents could receive multiple labels when they contained more than one type of sensitive-to-remember information. The researchers met regularly to resolve unclear boundaries, refine category definitions, and identify illustrative examples. During this process, \textit{Stigmatized or taboo activities} emerged as an additional category.
No new categories emerged after 96 participants had contributed 143 incidents. We nevertheless recruited 50 additional participants to examine whether subsequent incidents could be accommodated by the developing taxonomy. The resulting taxonomy of 15 sensitive-to-remember categories was then reviewed by the third researcher, who confirmed that the category definitions and boundaries adequately represented the collected incidents.

Using the finalized taxonomy, the two researchers independently coded the incidents and achieved strong inter-rater reliability ($\kappa=.83$). The final taxonomy is reported in Section~\ref{sec: taxonomy}, with illustrative examples in Table~\ref{tab:taxonomy-examples}. We do not report category frequencies as prevalence estimates because participants were recruited specifically for prior forget experiences and could contribute different numbers of incidents; the resulting distribution therefore does not represent the prevalence of these sensitive-to-remember categories in broader AI use.

\paragraph{Identifying user needs.}
In parallel with taxonomy development, we analyzed participants' incident descriptions and responses about their current management strategies and desired alternatives for recurring needs related to post-sharing memory control. Two researchers inductively coded these responses, compared their interpretations, and consolidated recurring concerns into four higher-level user needs. The resulting needs informed the design goals reported in Section~\ref{sec:user needs}.

\subsection{User Evaluation}

We conducted a within-subjects study comparing \fence, a system we developed that operationalized the resulting taxonomy as rules for gating sensitive-to-remember conversations from being stored in AI memory, with conventional post-hoc memory management for information participants did not want an AI assistant to remember. Each participant completed the same task in both conditions in counterbalanced order.

\subsubsection{Participants}

We recruited 15 participants who used AI services at least weekly and had previously experienced wanting information shared with an AI not to be retained in memory. Eight were recruited through Prolific and seven through social recruitment. A pre-study survey assessed eligibility and elicited an incident used to prepare the study task. Participants received \$15. The final sample included 9 men and 6 women, aged 18--64. Full demographics are reported in Table~\ref{tab:participant-demographics}. We conducted 4 pilots to refine the protocol; pilot participants were not included in the final sample.

\subsubsection{Study Setup}

Sessions were conducted remotely via Zoom and lasted approximately 60 minutes. Participants remotely controlled the researcher's shared screen to interact with two conditions in a locally hosted web app.

In the \textit{baseline condition} (C0), information could enter persistent memory, after which participants could inspect and manage remembered information. In the \textit{policy-gated condition} (C1), \fence retained the same post-hoc memory controls as C0 while adding user-selected taxonomy categories as standing protection rules. As part of the initial setup, participants selected the categories they wanted \fence to protect before beginning the task. Conversations containing protected information were prevented from contributing to persistent memory, and the interface indicated that they were ``kept out of the AI's memory.'' Section~\ref{sec:system design} describes the system and controls in detail.

Both conditions were initialized with 12 saved memory items and 8 previous conversations to simulate an established account. For each participant, we created a fictional scenario based on the incident reported in the pre-study survey, preserving the type of disclosure and concern without requiring participants to re-enter their own sensitive information. The same scenario and prepared prompt were used in both conditions. The full protocol is provided in Appendix~\ref{app:evaluation-protocol}.

\subsubsection{Procedure}

\paragraph{Introduction and warm-up.}
After obtaining informed consent and permission to record, we asked participants about their AI use and understanding of AI memory. They then elaborated on their pre-study incident, including what they had been trying to accomplish and why they did not want the information remembered.

\paragraph{Comparative tasks.}
Participants completed the same task in both conditions in counterbalanced order. In each condition, they received the prepared scenario and prompt and were asked to imagine needing the AI's help with the situation. In C1, participants first completed the category-selection setup before sending the prompt. After sending the prompt, they were asked to assume that they did not want the disclosed information retained over time. Participants could take any action they considered appropriate, explore the available controls, or take no action while thinking aloud. Researchers did not direct them toward any specific control or interaction.

After each condition, participants completed the same questionnaire. We measured usability with the four-item Usability Metric for User Experience (UMUX)~\cite{Finstad2010UMUX} and mental effort with Paas's single-item nine-point scale~\cite{Paas1992MentalEffort}. We assessed interpretability of memory behavior using four adapted items from the \emph{Understandability/Predictability} subscale of the Trust in Automation (TiA) questionnaire and trust using two adapted items from its \emph{Trust in Automation} subscale~\cite{Koerber2019Trust}. All adapted items referred specifically to memory-management behavior. The complete questionnaire is provided in Appendix~\ref{app:evaluation-questionnaire}.

\paragraph{Exploration and follow-up interview.}
After completing both comparative tasks and questionnaires, participants returned to C1 for further exploration and a semi-structured interview. We asked them to explain their earlier actions and category selections, used their questionnaire responses to probe differences between conditions, and discussed whether the taxonomy and controls matched their expectations. Participants were invited to try additional sensitive and non-sensitive conversations with \fence to explore cases beyond the prepared scenario. We concluded with questions about perceived protection, existing memory-management practices, and desired improvements, followed by a demographic survey. Because this additional C1 exploration occurred only after both comparative tasks and questionnaires were completed, it did not affect the between-condition comparison.

\subsubsection{Analysis}

We scored the post-condition questionnaire by reverse-coding negatively
worded items and aggregating items within usability, memory interpretability, and trust in memory governance. Mental effort was analyzed as the single-item rating. Because the study used a within-subjects design with a small sample, we compared participants' ratings between C0 and C1 using two-sided Wilcoxon signed-rank tests~\cite{Wilcoxon1945}. We applied a Bonferroni correction
across the four overall measures to control the family-wise error rate across
these comparisons~\cite{Dunn1961MultipleComparisons}.

We analyzed participants' transcripts and observation notes of their interactions during the tasks using thematic analysis.
We treated these behavioral observations as complementary evidence for interpreting participants' accounts, rather than as a separate behavioral dataset.
Two researchers independently open-coded materials from four participants to identify recurring themes and then compared their initial codes, resolved disagreements through discussion, and developed a codebook (Appendix~\ref{app:qualitative-codebook}). The researchers subsequently coded the remaining transcripts and notes, applied and refined the codebook, and identified broader themes for reporting. We assessed inter-rater reliability using Cohen's kappa for code application, yielding $\kappa=.87$.

\subsection{Limitations}

Our survey relied on incidents that participants could recall and already recognized as information they wanted forgotten, and may therefore miss less salient concerns or those that emerge only through later reuse. We mitigated this limitation by continuing recruitment after category saturation, adding 50 participants after no new categories emerged to further assess coverage within our sample. Future research could explore collecting such incidents directly from interaction logs where participant consent and data-access conditions permit.

Our user evaluation focused on participants' initial experience with \textsc{Fence} rather than sustained everyday use. We grounded the study in participant-informed scenarios and open-ended exploration while avoiding the need to re-enter actual sensitive disclosures. Longer-term deployment could examine how trust, protection preferences, and management effort evolve over time, as well as whether other enforcement granularities produce different tradeoffs.

Finally, both studies involved U.S.-based, English-speaking participants. Future work should examine how the taxonomy and protection design transfer across cultural contexts with different disclosure norms and privacy expectations.

%% file: sections/4_Taxonomy.tex
\section{Taxonomy of Sensitive-to-Remember Information for AI Memory}

\label{sec: taxonomy}

We derived 15 categories of sensitive-to-remember information that together constitute the negative space of AI memory. Table~\ref{tab: taxonomy} presents the resulting taxonomy and category descriptions.
We provide illustrative examples for each category in Appendix~\ref{app: taxonomy example}. We intentionally use a flat structure so that categories can serve as a shared, interpretable vocabulary between users and the system, without requiring users to navigate a hierarchical classification scheme.

\newcolumntype{L}{>{\raggedright\arraybackslash}X}

\begin{table*}[t]
\centering
\caption{Taxonomy of sensitive information for AI memory. 
Illustrative examples and boundary cases for each category are provided in
Appendix~\ref{app: taxonomy example}.}
\label{tab: taxonomy}
\small
\renewcommand{\arraystretch}{1.15}
\setlength{\tabcolsep}{6pt}

\begin{tabularx}{\textwidth}{
    @{}
    >{\raggedright\arraybackslash}p{0.025\textwidth}
    >{\raggedright\arraybackslash}p{0.32\textwidth}
    L
    @{}
}
\toprule
\# & \textbf{Category} & \textbf{Description} \\
\midrule

1 &
\textbf{Personal Identifying Information} &
Information that can identify, locate, contact, authenticate, or profile an
individual, either directly or in combination with other information. \\

2 &
\textbf{Health information} &
Explicit information concerning physical health or clinically meaningful mental-health
conditions, including symptoms, diagnoses, treatment, disability, or related
care-seeking. \\

3 &
\textbf{Financial information} &
Information concerning a person's finances, financial circumstances, or
financial activities. \\

4 &
\textbf{Legal and administrative proceedings} &
Information that discloses a person's involvement in legal, governmental, or administrative processes. \\

5 &
\textbf{Professional identity and career-sensitive information} &
Information that discloses a person's specific professional or academic identity, or information that could affect a person's career opportunities, standing, or reputation. \\

6 &
\textbf{Intimate body and sexual information} &
Information that discloses a person's sexual, reproductive, and intimate bodily information or experiences. \\

7 &
\textbf{Romantic and intimate relationships} &
Sensitive information about a person's romantic, intimate, or desired relationships and their dynamics. \\

8 &
\textbf{Family and close-relation information} &
Private information concerning a person's family members, dependents, household members, or other close relations. \\

9 &
\textbf{Risky interpersonal relationships} &
Information that discloses a person's sensitive dynamics or tensions within non-romantic, non-family social relationships. \\

10 &
\textbf{Emotionally vulnerable disclosure} &
Information revealing vulnerable emotional states or experiences, such as
shame, loneliness, fear, despair, humiliation, or emotional dependence. \\

11 &
\textbf{Negative self-image} &
Information expressing negative evaluations of oneself or perceived personal
inadequacy. \\

12 &
\textbf{Adverse life events and traumatic experiences} &
Experiences involving substantial harm, victimization, death, or other severe life events. \\

13 &
\textbf{Stigmatized or taboo activities} &
Information that discloses activities or behaviors that may carry substantial social stigma, moral judgment, or cultural taboos. \\

14 &
\textbf{Cultural and ethnic background} &
Information about identity, background, or lived experiences associated with
race, ethnicity, culture, ancestry, or heritage. \\

15 &
\textbf{Ideological beliefs} &
Information that discloses someone's personal beliefs or commitments related to religion, politics, or other ideologies. \\

\bottomrule
\end{tabularx}
\end{table*}

Alongside domains captured by prior information-sensitivity \cite{Milne2017InformationSensitivity} and self-disclosure taxonomies \cite{Dou2024SelfDisclosure}, our taxonomy surfaces forms of sensitive-to-remember information that are less directly represented in these frameworks, such as negative self-image, adverse life events, and risky interpersonal conflict. It also draws more selective boundaries within broad domains that prior taxonomies treat primarily as information types.

For example, professional information is not sensitive simply because it concerns work: a user may be comfortable with the AI remembering that they work in tech, while preferring it not to retain that they are currently looking for new opportunities while still employed because of the potential career or reputational consequences.
This distinction is particularly important for AI memory. As reflected in the user needs and threats discussed in Sections~\ref{sec:user needs} and~\ref{sec: treat model}, users' concerns arise not only from what information explicitly identifies them, but also from whether retained information remains linked to their profile, can be exposed to others, or can later shape inference and personalization. A taxonomy for AI memory therefore needs to capture both a broader range of personally consequential disclosures and the context in which otherwise ordinary information becomes sensitive to retain. 

In the following section, we use this taxonomy to inform our policy-level protection design. We also evaluate its operationalizability by examining human annotations produced using the taxonomy and evaluating model-generated labels against them; we report this technical evaluation in Section~\ref{sec:tech evaluation}.

%% file: sections/5_Design.tex
\section{A Design for Policy-level Protection}

Building on the taxonomy, we designed \fence as a policy-level protection layer that keeps user-selected classes of information from entering persistent AI memory. This section presents the design of \fence{}, including the user needs that motivated it, the design goals we derived, and how those goals were implemented in the system.

\subsection{User needs}
\label{sec:user needs}
Our survey revealed four recurring needs that shaped the design of \fence{}.

\paragraph{U1: Make memory status visible.}
Participants often did not know whether information had entered memory or remained within the current conversation. Some only realized afterward that a disclosure had been retained (e.g., I4, I82, I143), while others had not considered storage at the time of sharing (e.g., I5, I119, I168). Uncertainty also remained after participants tried to remove information: they questioned whether it was ``truly gone'' (I5), ``fully removed'' (I82), or wanted explicit confirmation that forgetting had occurred (I106, I107). These accounts point to a need for visible memory and protection states before and after management actions.

\paragraph{U2: More effortless and feasible ways of management.}
Participants described current management as difficult to locate and effortful to use. They struggled to search long memory lists, find the relevant conversation, or access deletion controls (I1, I12, I37, I55). Others reported doing nothing because they did not know what action would work or felt there was little they could do after disclosure (I17, I84, I92). Post-sharing protection therefore needs to be readily available without requiring users to locate and manage individual memories one by one.

\paragraph{U3: Provide assurance after disclosure.}
For many participants, the problem was not only finding a deletion mechanism but trusting that any action could undo the effects of disclosure. Participants described repeatedly deleting or asking the AI to forget while remaining unsure whether the information persisted (I10, I21, I53). Some worried about later reuse, while others were troubled by the information simply continuing to exist in the system, describing it as having been ``ingested by the model'' or becoming a persistent ``digital file'' (I57, I136). These accounts highlight a need for protection that gives users confidence that sensitive disclosures will not continue to shape persistent memory.

\paragraph{U4: Preserve a safe space for necessary disclosure.}
Avoiding sensitive disclosure altogether was not always realistic. Participants shared personal context to seek relationship advice, mental-health support, sexual-health information, or legal guidance (e.g., I20, I39, I61, I87). Others disclosed more than intended while distressed, sleep-deprived, intoxicated, or otherwise poorly positioned to self-censor (I3, I46, I50, I63). These cases suggest that users need to share sensitive context without automatically turning that disclosure into persistent memory.

\subsection{Threat model}
\label{sec: treat model}

We consider threats arising when sensitive information disclosed to an AI assistant is retained and linked to a user's persistent memory or profile. We identify three threats that may arise when users rely on remotely provided AI systems whose memory behavior may not align with their needs: the system may (1) retain sensitive-to-remember information, (2) reveal sensitive-to-remember information to unintended viewers, such as friends or others who gain access to the user's AI account, and (3) reuse sensitive-to-remember information in ways that lead to inappropriate inferences or unwanted personalization of future responses. We acknowledge broader provider-side privacy threats, including the use of data for model training, internal data access, and back-end breaches. These threats are outside our scope; we focus on risks arising from information retention within the user-facing AI memory environment.

\subsection{Design Goal}

We aim to design memory governance at the \textbf{\textit{policy level}}.
We characterize this protection through three properties. First, its object is system behavior rather than an individual memory item.
Second, its scope extends across a class of cases and across interactions.
Third, it establishes a standing system obligation rather than relying on repeated user intervention.
Making such rules actionable requires a way to determine which disclosures should be governed alike. We use the taxonomy introduced earlier as this policy vocabulary: users select the sensitive-to-remember categories they want protected, and the system applies the corresponding memory rule whenever a conversation falls within those categories. This follows a broader pattern in policy design, where classification structures determine which rules or protections apply~\cite{Yang2024FutureHCIPolicy,GDPR2016,EUAIAct2024,NISTSP80060}. The taxonomy therefore represents one operationalization of policy-level protection rather than a requirement of the broader design paradigm.

Guided by our overarching design goal, we specified four subgoals to ground our solution and address the user needs identified in Section \ref{sec:user needs}:

\paragraph{D1 Protection by default, not by choices}
Our findings suggest that sensitive information should be protected by default rather than relying on users to make the right privacy choice at the right moment. Participants were often unaware of what would be remembered, and sometimes disclosed sensitive information when such context was necessary to complete a task. Requiring users to anticipate these risks and manually configure protection therefore places too much responsibility on them (U2). Instead, protection should be built into the memory process itself, withholding potentially sensitive information from persistent memory unless there is a clear reason to retain it (U2, U3).

\paragraph{D2 Make memory decisions tangible and explainable}
Our findings show that users often struggled because memory remained an invisible and ambiguous system process. We therefore aim to make memory decisions tangible and interpretable by showing what information is treated as memory, whether it is retained, why it is saved or withheld, and how user actions change its state (U1). Beyond providing a generic memory setting, the system should make individual memory decisions visible and explainable, enabling users to understand what the system remembers about them and verify whether its behavior matches their expectations (U1, U2).

\paragraph{D3 Flexibility to support individual personalization needs}
Memory should remain flexible enough to support useful personalization. People use general-purpose AI across diverse personal and professional contexts, where remembering information about their roles, projects, and preferences can make future interactions more relevant and efficient~\cite{chatterji2025people}. What is useful or appropriate to remember therefore differs across users and contexts (U4). Our design should avoid treating any information category as universally useful or unwanted, allowing protection to adapt to individual preferences while preserving memory that users find valuable (U4).

\paragraph{D4 A trustworthy environment without hiding}
Many sensitive disclosures in our study arose from genuine needs for help, including personal advice, emotional support, and tasks requiring detailed context. Yet these interactions were often followed by regret, embarrassment, or anxiety about what had been shared. Our design should therefore provide not only functional privacy protection but also a trustworthy environment in which users can seek help without worrying that necessary disclosures will create lasting consequences (U3, U4). The goal is to make sensitive interactions safe enough to remain useful while giving users confidence that the system will handle their disclosures with care (U4).

\begin{figure}[t]
    \centering
    \includegraphics[width=\columnwidth]{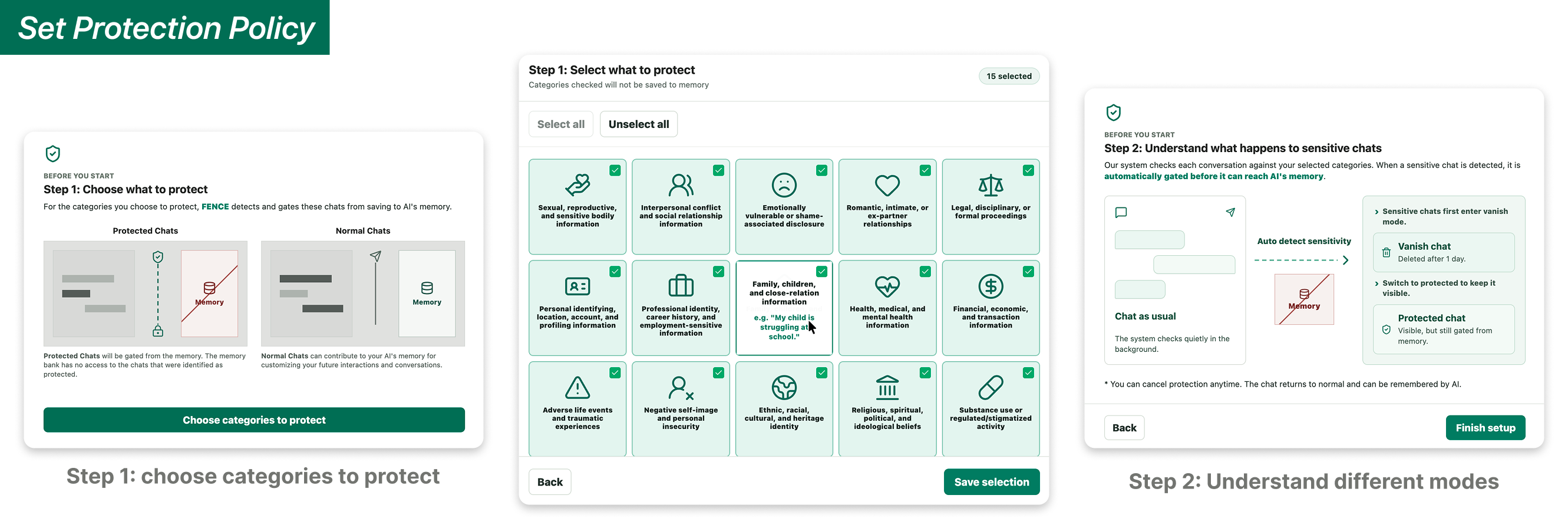}
    \caption{Setting a standing protection policy in \fence{}. During onboarding, users first select the sensitive-to-remember categories they want protected and then learn how protected and normal conversations differ in their memory behavior.}
    \label{fig:design-setup}
\end{figure}

\subsection{The FENCE System}
\label{sec:system design}

Guided by these design goals, we propose \fence{}, a memory protection layer that proactively gates potentially sensitive disclosures before they enter an AI assistant's memory. Figures~\ref{fig:design-setup} and~\ref{fig:design-use} show how users establish a standing protection policy and how \fence{} applies that policy during subsequent conversations. Consistent with our characterization of policy-level protection, \fence{} places user control primarily over the governing rule rather than individual memory items, providing protection by default for the user's selected negative memory space.

\subsubsection{Establishing a Standing Policy Before Disclosure}
\label{sec:setup}

\begin{figure}[t]
    \centering
    \includegraphics[width=\columnwidth]{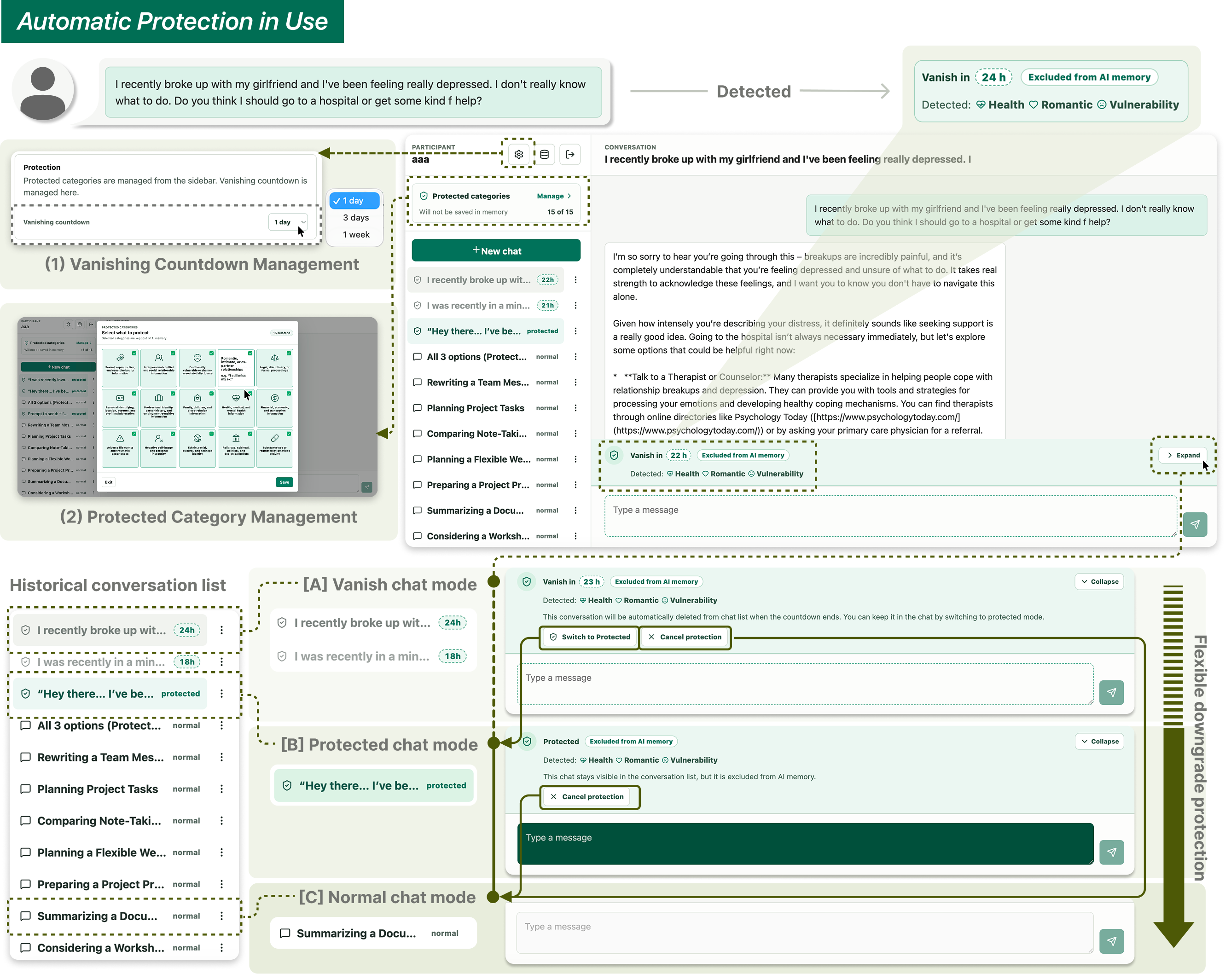}
    \caption{Automatic protection during use. \fence{} detects disclosures covered by the user's standing policy and applies protection automatically. Users can manage the vanishing countdown (1) and protected categories (2), while individual conversations can be relaxed from (A) \textit{Vanishing} to (B) \textit{Protected} or (C) \textit{Normal}.}
    \label{fig:design-use}
\end{figure}

\fence{} front-loads policy specification into a two-step onboarding flow, so that the governing rules exist before sensitive disclosures occur (Figure~\ref{fig:design-setup}). Users first specify the categories of information they want to protect. We instantiate the taxonomy introduced in Section~\ref{sec: taxonomy} as 15 configurable categories, each with an icon and a short example to support recognition rather than recall~\cite{Nielsen1994,habibTogglesDollarSigns2021}. The categories make the policy expressible by giving users a vocabulary for declaring which classes of disclosure should be treated alike, while later allowing \fence{} to explain why a conversation was gated (D2). All categories are initially selected for protection by default (D1), while users can reconfigure their selections to accommodate different personalization needs (D3)~\cite{acquistiNudgesPrivacySecurity2017}. Users can also revise protected categories later through the management interface shown in Figure~\ref{fig:design-use}(2).

The second onboarding step explains how protected and normal conversations differ in their visibility and memory behavior (Figure~\ref{fig:design-setup}), helping users form an accurate model of how memory is governed in \fence{} (D2, D4).

\subsubsection{Applying the Policy During Conversation}
\label{sec:auto}

Once configured, the policy applies without further user action. In Figure~\ref{fig:design-use}, a user discloses emotionally vulnerable information about a recent breakup and asks whether they should seek medical help. \fence{} identifies the disclosure as matching \textit{Emotionally Vulnerable}, \textit{Romantic}, and \textit{Health}, and automatically places the conversation in \textit{Vanishing} mode without requiring the user to recognize the sensitivity of the disclosure or intervene at the moment of sharing (D1, D4). We use the conversation as the unit of enforcement because it preserves the context needed for the current interaction while giving users a clear boundary for what will not enter persistent memory (D2, D4).

\textit{Vanishing} (Figure~\ref{fig:design-use}A) is the default protective state. The conversation is excluded from AI memory immediately and removed from the conversation list when the countdown expires. In our implementation, expiration also deletes the conversation from the application database. This design addresses threats (1) and (2) in Section~\ref{sec: treat model}, as participants expressed concerns about both persistent retention and continued visibility of sensitive information. Prior work on shared LLM accounts similarly finds that users delete sensitive queries, use temporary chats, or avoid private disclosures because co-users may view their histories~\cite{Song2026DontLook}. Users can configure the vanishing period through the control shown in Figure~\ref{fig:design-use}(1), with a 24-hour default that can be extended to three days or one week.

The \textit{Protected} state (Figure~\ref{fig:design-use}B) keeps the conversation visible and accessible in the conversation history while excluding it from persistent AI memory. Protected chat therefore separates conversation retention from memory contribution.

Users can relax protection on an individual conversation, moving from \textit{Vanishing} to \textit{Protected} or \textit{Normal}, or from \textit{Protected} to \textit{Normal} (Figure~\ref{fig:design-use}A--C). \fence{} initially applies the most protective state and lets users deliberately relax protection when retention or personalization becomes desirable. This preserves protection by default (D1) while supporting changing personalization needs (D3).

\subsubsection{Making Protection States Recognizable}
\label{sec:states}

\fence{} displays detected categories and the resulting protection state in a banner above the message input, keeping this information in the user's line of sight rather than in a separate settings interface~\cite{Muller2022Noticeability} (D2). Reporting the triggering categories rather than only the outcome helps users understand why protection was applied~\cite{Kulesza2013TooMuch} and assess whether the system's behavior matches their expectations, supporting calibrated trust~\cite{Kizilcec2016} (D2, D4). The banner can expand to explain what will happen to the conversation and provide controls for changing its state, as shown in Figure~\ref{fig:design-use}A--B.

Beyond these textual cues, \fence{} differentiates the three protection states visually in both the conversation area and conversation history (D2). In \textit{Vanishing} (Figure~\ref{fig:design-use}A), dashed elements communicate impermanence and the conversation list displays the remaining countdown, drawing on the visual language of existing temporary-chat interfaces~\cite{OpenAI2026TemporaryChat}. In \textit{Protected} (Figure~\ref{fig:design-use}B), a solid boundary communicates persistence while the protected styling indicates that the conversation remains excluded from memory. \textit{Normal} (Figure~\ref{fig:design-use}C) retains the familiar appearance and memory behavior of conventional AI chat, reducing the need to learn a separate interaction model~\cite{Nielsen1994}.

\subsubsection{Implementation}
\begin{figure*}[t]
    \centering\includegraphics[width=\textwidth,height=0.85\textheight,keepaspectratio]{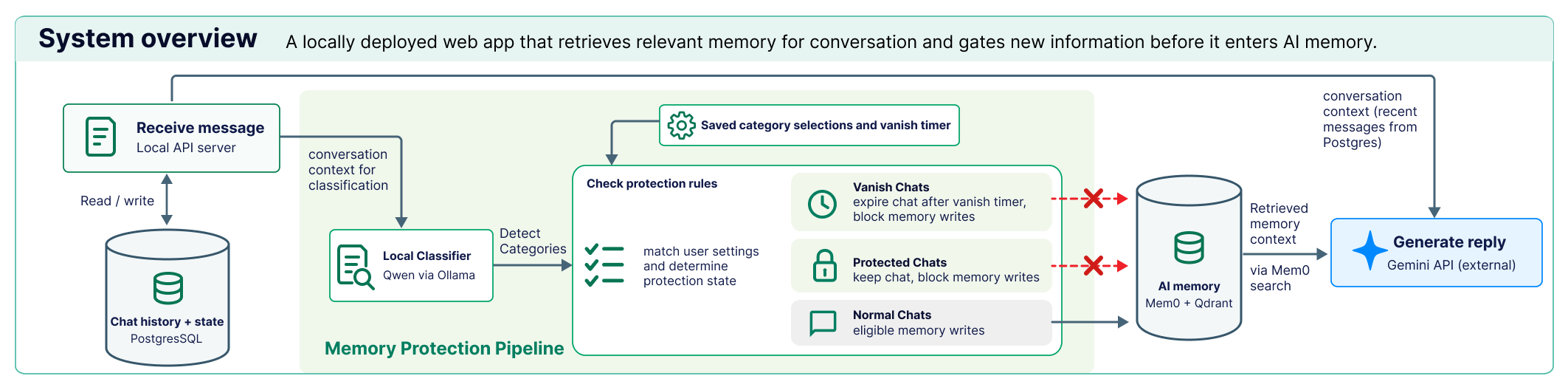}
    \caption{
    Overview of \textit{FENCE}'s system architecture and data flow. Incoming messages are stored in PostgreSQL and classified locally (Qwen via Ollama) to detect sensitive categories, which are checked against the user's saved protection settings to set the conversation's protection state. Normal chats remain eligible for storage in AI memory (Mem0 and Qdrant); protected and vanishing chats are excluded. Retrieved memory is combined with the ongoing conversation to generate a reply via the Gemini API.
    }
    \label{fig:system-overview}
    \Description{Overview of the FENCE system workflow
    }
\end{figure*}

As shown in Figure~\ref{fig:system-overview}, the locally deployed application uses Qwen3.5-9B via Ollama to classify disclosures and checks the detected categories against the user's protection rules. The resulting conversation state governs memory eligibility, while the Gemini API generates conversational responses. We used OpenAI's coding agent, Codex, to help develop the system, with one researcher inspecting the implementation details and two researchers testing the system's performance.

%% file: sections/5_Model_benchmark.tex
\begin{table}[t]
    \centering
    \small
    \setlength{\tabcolsep}{4.5pt}
    \renewcommand{\arraystretch}{1.08}
    \caption{Model performance across all 19,500 message--category
    decisions and by the level of agreement among the three researchers.
    \emph{Unanimous (3/3)} includes decisions for which all three
    researchers assigned the same binary label, whereas
    \emph{Majority (2/3)} includes split decisions for which exactly two
    researchers assigned the same label. Here, $n$ denotes the number of
    message--category decisions rather than messages. Balanced accuracy
    (BA) gives equal weight to recall for positive and negative labels.
    Across all four models, unanimous decisions produced higher BA and
    recall, whereas majority decisions produced higher precision under
    their greater positive-label prevalence. Underlining indicates the
    highest value within each model, and bold indicates the highest value
    overall.}
    \label{tab:performance-by-human-agreement}

    \begin{tabular}{llrrrr}
        \toprule
        \textbf{Model} &
        \textbf{Human agreement} &
        \textbf{$n$} &
        \textbf{BA} &
        \textbf{Recall} &
        \textbf{Precision} \\
        \midrule

        \multirow{3}{*}{Gemma 3}
        & All decisions
        & 19,500 & .838 & .796 & .439 \\
        & Unanimous 
        & 18,183 & \underline{.860} & \underline{.831} & .351 \\
        & Majority 
        & 1,317 & .635 & .746 & \underline{.735} \\
        \addlinespace

        \multirow{3}{*}{Qwen3.5-9B}
        & All decisions
        & 19,500 & .886 & .817 & .682 \\
        & Unanimous 
        & 18,183 & \underline{.928} & \underline{.893} & .630 \\
        & Majority 
        & 1,317 & .700 & .705 &
        \underline{\textbf{.804}} \\
        \addlinespace

        \multirow{3}{*}{Gemini 3.8 Flash}
        & All decisions
        & 19,500 & .944 & .944 & .664 \\
        & Unanimous 
        & 18,183 &
        \underline{\textbf{.971}} &
        \underline{\textbf{.984}} &
        .626 \\
        & Majority 
        & 1,317 & .660 & .886 & \underline{.735} \\
        \addlinespace

        \multirow{3}{*}{GPT-5.6 Luna}
        & All decisions
        & 19,500 & .935 & .931 & .644 \\
        & Unanimous 
        & 18,183 & \underline{.965} & \underline{.977} & .596 \\
        & Majority 
        & 1,317 & .664 & .863 & \underline{.741} \\

        \bottomrule
    \end{tabular}
\end{table}

\subsection{Benchmarking Human-Model Taxonomy Agreement}
\label{sec:tech evaluation}

We curated a benchmark of 1,300 messages to test whether language models can operationalize our taxonomy as a multilabel classification task. For each of the 15 categories, the benchmark contains 50 synthetic messages and 20 real user messages from WildChat~\cite{Zhao2024WildChat} and LMSYS-Chat-1M~\cite{zheng2023lmsyschat1m}. The benchmark also includes 250 non-sensitive messages to assess false positives. Three researchers assigned each message 15 binary labels (19,500 message--category decisions); the majority vote served as the reference label, and 93.2\% of decisions were unanimous. Importantly, these annotations capture whether a message falls within a taxonomy category, not whether a particular user would want that message protected. The benchmark therefore evaluates models' ability to apply the taxonomy semantically rather than predict individual privacy preferences.

We evaluated two locally deployable open-weight models, Gemma~3 4B and Qwen3.5-9B, and two frontier models, Gemini~3.8 Flash and GPT-5.6 Luna (see Appendix~\ref{app:model-classification-prompt} for the prompt). As only about one in ten decisions was positive, we report balanced accuracy (BA), recall, and precision, stratified by human agreement (Table~\ref{tab:performance-by-human-agreement}).

Gemini~3.8 Flash achieved the highest BA (.944) and recall (.944), closely followed by GPT-5.6 Luna (.935 and .931, respectively). Qwen3.5-9B trailed Gemini by less than .06 in BA (.886; within .05 on unanimous decisions) while achieving the highest precision (.682) and a comparable recall (.817). These results support the feasibility of applying the taxonomy with a locally deployable model, although a performance gap remains relative to frontier models. Future work could examine whether prompt calibration or fine-tuning can narrow this gap, particularly in recall.

Both frontier and local models showed higher recall than precision, an error profile compatible with D1's emphasis on protecting memory's negative space by default. In this setting, false positives extend protection to conversations that could otherwise contribute to personalization, whereas false negatives leave protected information eligible for memory. This tradeoff supports a conservative default, while also underscoring the cost of repeated over-protection for personalization.

All models performed best on unanimous decisions, with BA falling to .635--.700 on split decisions. The decline was particularly large for the strongest models (Gemini~3.8 Flash: .971 $\rightarrow$ .660; GPT-5.6 Luna: .965 $\rightarrow$ .664), suggesting that ambiguity in category boundaries contributes substantially to the remaining difficulty. For example, human agreement was lowest for \emph{Adverse life events and traumatic experiences}, where researchers disagreed over whether a severe workplace injury qualified.

This ambiguity also clarifies the role of classification in \fence{}. Our reference labels reflect majority judgments from three researchers and should therefore be understood as an operationalization of the taxonomy rather than a population-level ground truth. The classifier is not intended to infer a user's desired memory outcome for every message. Instead, it applies a standing boundary that the user has already chosen. A message may therefore be correctly classified while still warranting a user override; such cases reflect an exception to the standing policy rather than a classification error. Future work should validate category boundaries with broader annotator populations and user-specific judgments, and examine whether calibration or adaptive policies can reduce the need for overrides without returning to case-by-case preference prediction as the primary control mechanism.

%% file: sections/6_User_evaluation.tex
\section{User Evaluation Results}

Participants generally evaluated \textsc{Fence} more positively than baseline memory management on usability, interpretability, and trust, although it did not yield a significant reduction in mental effort. Qualitative findings show how a standing, category-based policy made protection more understandable and offered reassurance about sensitive disclosures. They also reveal that confidence in automated protection depended on whether its decisions matched users' intended boundaries and could be verified beyond the interface's stated protection status.

\subsection{Quantitative Result}
\label{sec:quantitative-result}
We compared participants' ratings of the baseline condition (C0) and the policy-gated condition (C1) using two-sided Wilcoxon signed-rank tests, with Bonferroni correction across the four overall measures. Given the small sample, we interpret these comparisons as within-participant evidence of perceived differences between the prototypes rather than as population-level estimates. Composite-level results are summarized in Table~\ref{tab:ratings}.

The largest and most consistent difference concerned participants' understanding of the system's memory behavior: 12 of 15 participants rated C1 as more interpretable than C0, and the median composite score increased from 3.5 to 4.5 (Mdn$_{\text{C0}}=3.5$, Mdn$_{\text{C1}}=4.5$; $Z=-2.70$, $p_{\mathrm{adj}}=.028$, $r=.75$). This improvement was most pronounced in participants' understanding of why information was or was not saved (Mdn$_{\text{C0}}=4$, Mdn$_{\text{C1}}=5$; $Z=-2.75$, $p_{\mathrm{adj}}=.024$, $r=.92$). Participants also rated C1 as more usable (Mdn$_{\text{C0}}=5.75$, Mdn$_{\text{C1}}=6.25$; $Z=-2.59$, $p_{\mathrm{adj}}=.038$, $r=.72$) and more trustworthy (Mdn$_{\text{C0}}=3$, Mdn$_{\text{C1}}=4$; $Z=-2.61$, $p_{\mathrm{adj}}=.036$, $r=.75$).

Evidence for reduced mental effort was less clear. Although the median rating decreased from 4 to 3 on Paas's 9-point scale, this difference did not reach significance after correction ($Z=-1.80$, $p_{\mathrm{adj}}=.288$, $r=.60$). Participants' interviews help explain this pattern: while gated memory reduced some of the burden of monitoring what the system might remember, it also introduced new considerations around configuring protection and deciding when and how to intervene. We examine these tradeoffs, along with how participants understood and negotiated control over AI memory, in the qualitative findings that follow.

\begin{table}[t]
\centering
\caption{Composite-level ratings comparing the baseline condition (C0)
and the gated-memory condition (C1) across the four overall measures.
Wilcoxon signed-rank tests, Bonferroni-corrected across the four
measures; $N=15$. ``Higher in C1 / Tied / Higher in C0'' reports the
number of participants whose composite score was higher under each
condition or unchanged. Full item-level statistics are reported in
Table~\ref{tab:ratings-items} in Appendix~\ref{app:item-level-stats}.}
\label{tab:ratings}
\begin{tabular}{lccccccc}
\toprule
\textbf{Measure} & \textbf{Mdn (C0)} & \textbf{Mdn (C1)} & \textbf{Higher C1 / Tied / Higher C0} & \textbf{$Z$} & \textbf{$p_{\mathrm{adj}}$} & \textbf{$r$} \\
\midrule
Usability         & 5.75 & 6.25 & 10 / 2 / 3 & $-2.59$ & $.038$ & .72 \\
Trust             & 3.00 & 4.00 & 10 / 3 / 2 & $-2.61$ & $.036$ & .75 \\
Interpretability  & 3.50 & 4.50 & 12 / 2 / 1 & $-2.70$ & $.028$ & .75 \\
Mental Effort     & 4.00 & 3.00 & 2 / 6 / 7  & $-1.80$ & $.288$ & .60 \\
\bottomrule
\end{tabular}
\end{table}

\subsection{Qualitative Result}

\subsubsection{Current memory management left users uncertain about forgetting}

Baseline behavior echoed the transparency and management difficulties identified in our formative survey (U1, U2). Only three of 15 participants (P1, P2, P6) independently located the memory-management interface, despite being asked to manage information they did not want remembered. Seven instead used the more immediately available conversation interface to ask the AI to forget, delete, or stop retaining the information (P3, P4, P7, P9, P10, P12, P15), while two deleted the conversation itself (P5, P8). Others used different conversational strategies or took no clearly classifiable action.

Participants also did not treat deleting a conversation, deleting a memory item, and asking the AI to forget as equivalent. P1, for example, viewed memory items as ``inferences made from [the] chat'' rather than a complete account of what the AI might retain. They therefore deleted both the associated memories and the conversation to ``cover my bases,'' yet still remained unsure whether the AI could use its content. Similar uncertainty prompted others to layer or verify actions: P2 deleted both the memory item and conversation, P15 sought confirmation after asking the AI to forget, and P10 explicitly tested whether the AI could still recall the supposedly forgotten information.

\subsubsection{Default protection created a safer space for disclosure}

\textit{\textbf{Users preferred protection before memory formation.}}
Participants consistently distinguished \fence's upfront protection from existing approaches that required them to remove information after it had already entered the system. P1 described the default policy as more trustworthy because the AI recognized ``from the get-go'' that selected information should not be remembered, rather than requiring later cleanup. P8 similarly highlighted that the mechanism intervened ``before it even gets to the memory,'' viewing this timing as a core advantage of the design. For participants who were uncertain whether deletion could fully undo memory formation, preventing selected information from entering memory changed the interaction from remediation to prevention. The value of the mechanism therefore came not only from providing another management interface, but from shifting when privacy management occurred.

\textit{\textbf{Default protection reduced repeated cleanup over time.}}
A standing protection policy also reduced the need to revisit individual conversations as memory accumulated. Participants contrasted selecting protected categories once with repeatedly locating sensitive chats, deleting memory items, or issuing forgetting requests. P1 described the preset categories as making management ``a lot faster,'' while P3 and P6 similarly noted the burden of reviewing and cleaning up memory over long-term use. This advantage became particularly salient with frequent AI use, as previously stated preferences could be applied to future disclosures without requiring a new privacy decision each time.

\textit{\textbf{Automatic protection reduced anxiety around sensitive disclosure.}}
Beyond reducing effort, participants described automatic protection in explicitly affective terms. P15 said the protected condition felt ``much safer'' and made them feel ``so much comfortable'' because they knew protected information would not be carried into memory and the system could still intervene if they forgot to activate protection themselves. This reassurance mattered because participants could not always anticipate sensitive disclosure in the moment. P8 described sharing personal situations ``in the heat of the moment,'' while P1 recalled late-night conversations in which they were ``half awake'' and would ``ramble'' without fully thinking through their responses.
Similar responses described automatic protection as providing a sense of security or relief from having to remember a separate management action (P2, P13).

\subsubsection{Taxonomy translated privacy judgments into usable memory rules}

\textit{\textbf{Taxonomy labels explained protection decisions.}}
The taxonomy gave automated protection a human-readable explanation by connecting each system action to a recognizable type of information. P8 valued feedback stating that a conversation had been ``flagged as'' a particular category and explaining the resulting memory behavior, rather than leaving users to infer why protection occurred. Category specificity also helped users evaluate whether the system's interpretation matched their own; P15 noted that sufficiently specific categories reduced the risk of keeping information they wanted removed or removing information they wanted retained. In this way, the taxonomy translated automated detection into an interpretable rationale for memory decisions.

\textit{\textbf{Category rules made future handling predictable.}}
Once users understood how selected categories mapped to protection, they could form expectations about how similar disclosures would be handled later. P10 described being clear about ``the policies, the expectations that this platform provided''; when sensitive information was subsequently detected and protected as expected, they felt ``there's no point in me following up further.'' This contrasted with P8's experience of existing AI memory, which would ``just kind of do what it wants to do,'' sometimes forgetting desired information while retaining unwanted information. P8 instead understood \fence as establishing a ``default for everything that you set,'' while allowing individual chats to be adjusted when needed. Categories thus provided a stable policy language connecting current protection decisions to expectations about future system behavior.

\textit{\textbf{Taxonomy prompted awareness of sensitive information.}}
Reviewing the taxonomy prompted participants to reconsider what information warranted attention when interacting with AI. P3 described it as ``a good way for me to think about what content I should be more careful about'' and what should be removed from memory. P2 went further, noting that users may not realize that something they type is sensitive; automatically identifying its category could make them aware of that sensitivity and more cautious in later interactions. This awareness also operated in context rather than only during initial configuration. When \fence surfaced categories alongside an ongoing conversation, multiple participants expressed that it gave them ``a very clear idea'' (P6) of what sensitive information they had disclosed within the current session (P2, P3, P8, P10). The taxonomy therefore not only represented users' existing preferences, but also provided a vocabulary for recognizing privacy-relevant disclosures.

Reviewing the taxonomy prompted participants to reconsider what information warranted attention when interacting with AI. P3 described it as ``a good way for me to think about what content I should be more careful about'' and what should be removed from memory. P2 went further, noting that users may not realize that something they type is sensitive; automatically identifying its category could make them aware of that sensitivity and more cautious in later interactions. This awareness also operated in context rather than only during initial configuration. When \textsc{Fence} surfaced categories alongside an ongoing conversation, multiple participants (P2, P3, P6, P8, P10) reported that it gave them, in P6's words, ``a very clear idea'' of what sensitive information they had disclosed within the current session. The taxonomy therefore not only represented users' existing preferences, but also provided a vocabulary for recognizing privacy-relevant disclosures.

\textit{\textbf{Protection choices reflected users' models of AI use.}}
Participants did not select categories according to a uniform notion of sensitivity, but instead according to how they conceptualized AI and expected to use it. P9 compared AI to ``a stranger, except it knows everything,'' and therefore considered information such as race or religion less concerning when it was information a stranger could readily know. In contrast, P15 wanted an ongoing assistant to ``get to know me a little bit'' while still maintaining boundaries around more sensitive information, whereas P6 treated some emotional disclosures like diary entries, appropriate to share in the moment but undesirable for the AI to reference later.

How participants used AI further shaped these judgments. Those relying on AI for ongoing projects or personalized assistance valued persistent context because it reduced repetition. P4, for example, chose to retain professional background so it would not need to be re-explained in future chats. By contrast, episodic uses offered less benefit from long-term memory. Participants also anticipated different risks from persistent memory: some focused on unwanted personalization or resurfacing, while others were more concerned about provider-side retention or interpersonal exposure. These differences help explain why the same category could lead to different protection choices across users. Rather than defining what is universally sensitive, the taxonomy provided a common policy vocabulary that users could adapt to their own expectations and concerns.

\subsubsection{Users assessed reliability through alignment with their protection judgments}

\textit{\textbf{Repeated agreement with users’ judgments strengthened perceived reliability.}} Users are not expected to trust automated protection immediately; instead, participants described confidence as developing as the system repeatedly made decisions that matched their expectations. P9 explained that ``as long as it's continuously proven that it works,'' trust would increase over time, while P6 similarly expected confidence to grow as detection became more accurate. Reliability was therefore learned through repeated alignment with users' protection judgments rather than a single successful detection.

\textbf{\textit{Category boundaries shaped whether automated judgments matched users’ intent.}} Even when information technically fit a sensitive category, participants did not always believe protection was appropriate in that context. P8 illustrated this with fictional writing: a character’s traumatic history could be correctly recognized as sensitive content while still being information the user wanted the AI to retain for future story development. P10 raised a related distinction between information about the user and personal information about someone else, noting that a category labeled “personal identifying” could encourage users to interpret it only from a first-person perspective. Participants therefore valued enough taxonomy granularity to preserve distinctions that changed the desired memory outcome; as P15 noted, greater specificity could help avoid both ``delete[ing] stuff that you want to keep'' and retaining information users wanted excluded.

\textbf{\textit{Protection scope also shaped whether a correct detection produced the intended outcome.}} Participants sometimes agreed that a message contained sensitive information while disagreeing that the entire conversation should therefore be removed from memory. P8 noted that long-running chats could contain months or years of unrelated context, making whole-conversation exclusion “not ideal” even when one portion should be protected (P8). P6 similarly proposed redacting specific personally identifying information while retaining the rest of the conversation, separating interface-level or information-level protection from conversation-level memory exclusion. Reliability therefore reflected not only whether the system detected sensitivity correctly, but whether that judgment was applied at a scope consistent with what the user actually wanted protected.

\subsubsection{Users verified protection across multiple system surfaces}

\textbf{\textit{Visible protection states provided an initial basis for verification.}} Unlike baseline controls, where users often had to infer whether an action affected memory, \textit{FENCE} made its claimed protection state explicit. P2 contrasted this with manually deleting information, where even after deleting twice there was ``no notification'' confirming whether the memory was gone; seeing a protected state instead gave them ``some kind of security.'' P10 similarly became more willing to continue using \textit{FENCE} after seeing that sensitive information was automatically detected and protected. These signals did not prove what happened internally, but made the system's claimed memory state visible.

\textbf{\textit{Participants also verified protection across interfaces.}} Some participants looked for consistency between the conversation, memory interface, and later AI behavior. P5 described first “challeng[ing]” the AI directly by asking what it remembered, then checking the memory bank if further confirmation was needed (P5). P4 proposed a similar but more systematic comparison: checking whether a protected conversation appeared in the memory list while ordinary conversations did, and using that difference as evidence that \textit{FENCE} was storing them differently (P4). This suggests that verifiability came not only from making a protection decision visible, but also from giving users multiple representations whose agreement could provide stronger evidence about the resulting memory state.

\textbf{\textit{Later conversations became behavioral tests when visible states alone remained insufficient.}} P4 proposed opening a new chat and asking about both an ordinary conversation and a protected one: if the AI recalled the former but not the latter, they would “confirm” that the protected information had not been memorized. P10 enacted a similar strategy in the baseline condition. After asking the AI to forget a phone number and receiving confirmation, they immediately responded, “let me try. Let me test it,” and asked the AI to recall the number. Yet even when the AI claimed not to remember, the original information remained visible in the chat, leading P10 to say, “I don't believe it.” Such conflicting signals prompted further checking because a conversational assurance, a visible chat history, and the AI's retrieval behavior could imply different memory states. Users therefore treated verification as an active process of checking whether these signals remained consistent.

%% file: sections/7_Discussion.tex
\section{Discussion}

\subsection{The Negative Space of AI Memory}

\subsubsection{What is it?}
Our primary empirical contribution is to show that AI memory contains a substantial \emph{negative space}: information that users need or choose to disclose for an immediate interaction but do not want retained for future interactions. This space is not well captured by the privacy--utility tradeoff that has shaped much of the privacy-preserving literature on conversational AI, in which the objective is typically to minimize disclosure while preserving necessary information~\cite{Ngong2025ContextualPrivacy,zhou2026operationalizing,Zhou2025Rescriber}. The negative space exposes the limitation of this remember-first paradigm: information may be useful or necessary for the immediate interaction without thereby warranting persistent retention. For such information, protection should therefore be enforced at the point of memory formation rather than through retrospective deletion.
Two results support this claim. First, the negative space observed in our study was broad. The 179 forget incidents encompassed 15 distinct information categories, extending beyond conventional identifying and sensitive attributes to include disclosures such as emotional vulnerability, negative self-image, interpersonal conflict, and adverse life experiences. During our evaluation, four participants selected all 15 categories for protection by default. We do not interpret this as evidence that every instance in these categories should be universally forgotten. Rather, it shows that the information users may wish to exclude from AI memory extends well beyond narrowly defined PII or a small set of conventionally sensitive attributes~\cite{Mireshghallah2024TrustNoBot,Dou2024SelfDisclosure}.
Second, protecting this space did not produce the utility penalty that a conventional privacy--utility tradeoff would predict. In the gated condition, \fence excluded protected conversations from memory, yet participants reported stronger agreement that the system's capabilities met their requirements and lower levels of frustration. This shows that within the situations represented in our study, participants did not experience stronger protection as a loss of useful system capability. This asymmetry is what makes protection by default defensible for the negative space, even if it would be inappropriate as a general rule for all AI memory.

\subsubsection{Where should the line be drawn?} Recent industry developments make the difficulty of drawing this boundary particularly visible. In August 2026, Anthropic introduced a sensitive-topic setting for Claude memory that excludes topics such as health, race and ethnicity, religious beliefs, politics, and gender identity, while placing stronger restrictions on information such as government identifiers and financial account numbers~\cite{Anthropic2026MemoryEverywhere}. This move reflects the premise underlying \fence: some information should be treated differently \emph{at the point of memory formation}, rather than relying only on users to remove it later. However, the categories explicitly named in Claude's public documentation cover only part of the negative space surfaced in our study. Categories such as emotional vulnerability disclosure, negative self-image, risky interpersonal conflict, adverse life events, and career-sensitive circumstances are not explicitly articulated. Other platforms draw the boundary differently. OpenAI states that information considered useful for future conversations may be saved automatically when memory is enabled and that sensitive information may also appear in memory, while recommending Temporary Chat when users do not want a conversation to contribute to personalization~\cite{OpenAI2026MemoryFAQ}. These differences do not establish that any one provider has identified the appropriate boundary. Instead, they show that AI companies are already making policy decisions about the negative space, but those decisions remain provider-defined and can change as memory products evolve. Users consequently encounter different rules about what is remembered depending on the system they use, rather than a standardized and commonly understood protection boundary.

\subsubsection{What should happen within the space?} Once a negative space is established, our findings also suggest that protection within it should be conservative by default. Participants strongly valued FENCE's ability to enforce previously selected protection rules without requiring them to identify and correct each sensitive disclosure retrospectively. Several participants were willing to tolerate some over-protection within categories they had deliberately selected because information that was unnecessarily excluded could be added back to memory, whereas information that was unexpectedly retained had already crossed the boundary they were trying to protect. This suggests an asymmetry between the two errors. For information a user has designated as part of their negative space, failing to retain something potentially useful can often be corrected; unwanted memory formation is harder to meaningfully undo and, as our baseline condition showed, may itself create uncertainty and distrust. This asymmetry echoes a principle proposed for agentic AI: actions that can only be reviewed after the fact should be easier to reverse than those that require prior approval~\cite{shavit2023practices}. Current memory formation violates this condition, since it is typically reviewed only afterward yet is difficult to undo.
Memory systems should therefore not equate potential usefulness with permission to remember. A practical implementation could treat a user's protected categories as a standing \emph{memory admission policy}: matching information is withheld by default, while the interface provides a low-cost way to explicitly admit a useful exception. Such a design preserves personalization without making ``remember unless later deleted'' the implicit rule.

Protection needs to be verifiable rather than merely asserted. Participants in both conditions repeatedly sought evidence that memory management had actually taken effect. They inspected memory interfaces, questioned the AI about what it knew, and initiated later interactions to test whether supposedly forgotten information could still be recalled. Similar concerns have been raised in prior work showing that natural-language privacy commands such as ``forget this'' can be easy to issue while leaving users uncertain about how the command was interpreted or which underlying data state it changed~\cite{Li2026PrivacyControl}. Our findings therefore suggest that policy-level memory governance should produce an inspectable record of enforcement. Such a record should come from the system, not the model. Model-generated explanations can be plausible yet unfaithful to the process that actually produced a behavior~\cite{turpin2023language,lanham2023measuring}, and governance work on agentic AI accordingly cautions against relying on a model to retroactively justify its actions~\cite{shavit2023practices}. P10's disbelief when the AI claimed to have forgotten illustrates the same gap from the user's side.
Under such governance, a protected disclosure could expose the source content that triggered the rule, the policy category that applied, and the resulting memory state. When persistent knowledge is derived from multiple conversations rather than stored as a one-to-one memory item, systems should still expose enough provenance for users to determine where remembered information came from and what must change for it to stop influencing future interactions. This requirement becomes increasingly important as memory draws on distributed information sources. Gemini, for example, instructs users that they may need both to delete relevant chats and disconnect the underlying app to remove information originating from a Connected App, because either source can continue making the information available~\cite{Google2026GeminiMemory}. Rather than forcing users to reconstruct such dependencies themselves, memory systems should make the relationship between source information, persistent state, and policy enforcement inspectable. This echoes prior work arguing that users should be able to validate AI memory behavior independently of the model's generated response~\cite{Vaithilingam2025SemanticCommit}.

Finally, our study shows that defining a negative space does not imply that every detected category should always produce the same protection outcome. Our study shows that participants sometimes agreed with \fence's classification while disagreeing with the scope of the resulting action. For example, participants may generally prefer health information not to be retained but still want the AI to explicitly remember a specific piece of health information, such as information about an allergy. Such cases distinguish two decisions that memory systems may otherwise conflate: recognizing that information belongs to a privacy-relevant category and deciding what memory consequence should follow. A taxonomy can support the former without fully determining the latter. We therefore envision users not simply as recipients of a provider-defined sensitive-content policy, but as \emph{rule builders} who can establish standing category preferences and revise them when the context requires an exception. At the implementation level, this suggests combining persistent category-level rules with local overrides, while supporting information-level protection when whole-conversation exclusion would unnecessarily remove useful context. The policy remains stable enough to protect users without repeated intervention, but its boundary remains revisable as their use of AI changes.

\subsection{Implications for Design and Policy}

\paragraph{Memory governance for persistent AI agents.} Emerging agent-governance frameworks address how authority is delegated to agents, how their access and actions are bounded, and how autonomous behavior remains accountable \cite{South2025Delegation,OWASPAgentic2025,NISTAgentStandards2026}. More broadly, governance defines protection objectives, allocates decision-making authority, and establishes mechanisms for policy enforcement and oversight \cite{NISTAIRMF2023,South2025Delegation}. Our work extends this governance perspective to AI memory, beyond treating memory as either a technical operation or an object of user-facing control. First, alongside existing goals of memory utility, integrity, and security \cite{Wang2025MemoryPrivacy,Microsoft2026Memory,Pulipaka2026Poisoning}, our findings establish protecting memory's negative space as an explicit governance objective, making non-retention a prospective policy obligation rather than merely a corrective action. Second, we argue for extending policy-setting authority to users, rather than limiting their participation to reviewing and correcting provider-generated memories. \fence operationalizes this through standing, user-selected category rules, with the taxonomy providing a shared vocabulary that translates personal boundaries into system-enforceable memory admission policies. Our evaluation illustrates the value of this approach: only three of fifteen participants independently located conventional memory controls, whereas twelve rated \fence's memory behavior as more interpretable, with participants particularly valuing the predictability of future memory decisions. This becomes increasingly important as personal agents like Muse reach more users, retaining personal information over time and gaining authorization to act across connected services \cite{Meta2026Muse,Meta2026MuseDesign}. As accumulated memories increasingly inform agents' decisions across contexts and over time, relying on users to continuously review individual memories risks making human oversight difficult to sustain \cite{Baum2026Anticipatory,Mitchell2026Oversight}. These developments call for durable, user-authorized memory policies whose enforcement remains consistent across evolving interfaces and expanding information sources.

\paragraph{Standardize the memory boundary rather than individual interfaces.}
These findings motivate a more explicit policy framework for AI memory. Current memory products are evolving rapidly, and the meaning and location of memory controls can change substantially as providers redesign their systems. Prior walkthrough research already documents differences across conversational AI platforms in how users access, edit, delete, and constrain interaction-derived data, as well as ambiguity around what natural-language control actually changes~\cite{Li2026PrivacyControl}. Recent product changes further illustrate this instability: in August 2026, Claude made memory shared across chat and Cowork, exposed remembered information as editable topics, and introduced default sensitive-topic protection~\cite{Anthropic2026MemoryEverywhere}. A policy framework should therefore not depend on users continually learning each provider's latest interface. We instead argue for standardizing the \emph{governance of the transition into memory}: systems should distinguish information used for the present interaction from information incorporated into persistent user state, disclose the default rule governing that transition, and provide users with a durable way to modify it. Providers should also make material changes to these rules visible rather than silently expanding what is eligible for persistence. Such a framework would complement rather than replace existing data-protection principles. The GDPR already establishes data minimization, data protection by default, and a right to erasure~\cite{GDPR2016,EDPB2020DataProtectionDefault}; these principles provide a strong normative basis for limiting unnecessary persistence, but they do not specify how a user-facing AI memory layer should represent or govern the conversion of interaction information into persistent context. Likewise, general AI transparency requirements do not themselves establish what an AI assistant should be permitted to remember~\cite{EUAIAct2024}. Memory formation deserves to be made an explicit policy object.

\paragraph{Extend protection beyond conversational input.}
The need for such a boundary becomes more pressing as AI systems move beyond remembering what users type. Contemporary assistants can increasingly retrieve information from other services and act across them. Gemini's Personal Intelligence, for example, can draw on emails, files, calendar events, photos, contacts, location information, and inferences derived across connected sources~\cite{Google2026PersonalIntelligence}. Claude similarly shares memory between chat and Cowork, allowing information learned in conversational interaction to carry into an environment designed for broader task execution~\cite{Anthropic2026MemoryEverywhere}. These developments expand the negative space from a question about conversational disclosure into a question about what information from a user's broader digital environment is allowed to become persistent AI state. A memory-protection mechanism should therefore sit at the \emph{memory boundary}, rather than only at the text-input interface. Information retrieved from a document or connected service should be evaluated under the same admission policy as information directly typed by the user. Systems should also preserve provenance when information crosses this boundary, so that a remembered item remains connected to the source from which it was derived and can be reevaluated if that source or its policy changes. Otherwise, protecting sensitive conversational disclosure while allowing the same information to be reconstructed and remembered from connected sources provides little meaningful protection.

\paragraph{Use taxonomy as legible policy infrastructure.}
Our evaluation provides evidence that a taxonomy can serve as a useful medium for implementing such policy. Participants repeatedly used category labels to understand why \fence had acted, predict how it would behave in future cases, and decide whether its rule matched their own preference. A taxonomy therefore creates a shared vocabulary through which system behavior can be both machine-enforceable and legible to users. We do not, however, argue that our 15 categories should become a fixed regulatory list. The boundary cases in our evaluation demonstrate why a policy taxonomy requires continued empirical refinement: categories need enough specificity to support meaningful rules, while sensitivity and desired retention remain contextual. A stronger policy implementation would consequently make not only category names but their operational boundaries inspectable. Providers adopting category-based protection could publish definitions and representative boundary cases for each protected class, report how reliably their systems identify those classes, and expose the category-level rationale when a memory decision is made. Policy revisions should likewise be versioned so that users and auditors can determine when the governing boundary has changed. These mechanisms would make the policy inspectable rather than leaving ``sensitive information'' as an unspecified internal standard. Future research can build on our taxonomy to determine which categories warrant broadly shared default protection and where finer distinctions materially change users' desired memory outcomes.

%% file: sections/appendix.tex
\appendix

\section{Formative Survey Questionnaire}
\label{app:formative-survey}

The questionnaire is reproduced below with minor grammatical corrections. Consent and administrative fields are omitted. One incident was required, and the same questions were repeated for up to two optional additional incidents. Participants who reported less than one month of AI use or no experience wanting AI to forget disclosed information were screened out before the incident questions.

\subsection{AI Usage}

Participants first reported their use of AI conversational agents. Their responses were displayed before the recall question to help them revisit their experiences.

\begin{enumerate}
    \item What AI conversational agents have you used? Check all that apply.

    \textit{Options:} ChatGPT; Gemini; Claude; DeepSeek; Grok; Other; I have never used an AI conversational agent.

    \item Overall, how long have you been using AI conversational agent(s)?

    \textit{Options:} Less than a month; 1 to 6 months; 6 to 12 months; more than a year.

    \item Which of the following have you used the AI chatbot for? Please thoroughly examine your history and check all that apply.

    \textit{Options:}
    \begin{itemize}
        \item Edit or critique provided text
        \item Personal writing or communication
        \item Translation
        \item Argument or summary generation
        \item Write fiction
        \item Ask for how-to advice
        \item Tutoring or teaching
        \item Creative ideation
        \item Ask for health, fitness, beauty, or self-care advice
        \item Mathematical calculation
        \item Data analysis
        \item Computer programming
        \item Create an image
        \item Analyze an image
        \item Generate or retrieve other media
        \item Seek specific information
        \item Purchasable products
        \item Cooking and recipes
        \item Greetings and chitchat
        \item Relationships and personal reflection
        \item Games and role play
        \item Other
    \end{itemize}
\end{enumerate}

\subsection{Forget Incidents}

Participants were shown a summary of their selected platforms, usage duration, and activities, followed by the recall prompt below. The incident questions were presented as open-ended text fields.

\noindent Now recall your experiences:

\begin{enumerate}
    \item Have you ever disclosed some information but later wished the AI would forget about it? It could be about a concrete text/image/voice input, or a higher-level intention, identity, characteristics, etc.

    \textit{Options:} Yes; No.
\end{enumerate}

\noindent What kind of information have you ever disclosed but wished the AI would forget?

\noindent\textit{You are encouraged to report up to 3 instances, but 1 is required for a successful submission.}

\begin{enumerate}
    \item[a)] In what context did you share the info with AI?
    \item[b)] At that time, how did you expect AI to handle or remember this info?
    \item[c)] What is the information ``to-be-forgotten''? (Describe it without revealing the exact content.)
    \item[d)] At what stage did you realize that you wanted AI to forget this?
    \item[e)] What concerns do you have if AI remembers this information?
    \item[f)] What have you done to address your concern, if any? (e.g., delete a conversation, use temporary chat)
    \item[g)] What opportunities or obstacles did you face when applying your strategy in question f?
    \item[h)] If you were the designer, how would you design the AI to better handle this instance?
\end{enumerate}

\noindent Besides the above case, do you have another experience where you wanted information to be forgotten by your AI? Share it here. (Optional)

\noindent\textit{The same questions (a--h) were provided for Instances 2 and 3.}

\subsection{Closing Questions}

Participants then answered a question about their understanding of AI retention and could provide additional comments.

\begin{enumerate}
    \item To your knowledge, do any AI conversational agents you use retain or learn from your past conversations?

    \textit{Options:} Yes; No; Not sure.

    \item We welcome any additional thoughts about this topic or this study. Feel free to share them below. (Optional)
\end{enumerate}

\section{Illustrative Examples for the Sensitivity Taxonomy}
\label{app: taxonomy example}

Table~\ref{tab:taxonomy-examples} provides illustrative examples for each
category in our sensitivity taxonomy. These examples demonstrate the types of
information covered by each category and are not intended to be exhaustive.

\begin{table*}[t]
\centering
\caption{Illustrative examples for each category in our taxonomy of sensitive
information. Examples are illustrative rather than exhaustive.}
\label{tab:taxonomy-examples}
\footnotesize
\renewcommand{\arraystretch}{1.13}
\setlength{\tabcolsep}{6pt}

\begin{tabularx}{\textwidth}{
    @{}
    >{\raggedright\arraybackslash}p{0.025\textwidth}
    >{\raggedright\arraybackslash}p{0.31\textwidth}
    >{\raggedright\arraybackslash}X
    @{}
}
\toprule
\# & \textbf{Category} & \textbf{Illustrative Examples} \\
\midrule

1 &
\textbf{Personal Identifying Information} &
Names; phone numbers; email addresses; physical addresses; location details;
usernames; account information; school or workplace names; strong demographic
signals; IP or network clues. \\

2 &
\textbf{Health information} &
Symptoms; diagnoses; medications; therapy; disability; clinically meaningful
mental-health distress; medical records; treatment; insurance or medical
billing; health-related or mental-health-related care-seeking. \\

3 &
\textbf{Financial information} &
Income; debt; loans; banking; credit cards; taxes; purchases; rent; benefits;
financial hardship; family finances; financial conflict. \\

4 &
\textbf{Legal and administrative proceedings} &
Court or police matters; tax or government processes; immigration;
institutional complaints; disciplinary procedures; lawsuits; custody or divorce
proceedings; formal investigations. \\

5 &
\textbf{Professional identity and career-sensitive information} &
Employer or manager information; job searches; performance issues; workplace
conflict; professional reputation; confidential work context; career or
academic history. \\

6 &
\textbf{Intimate body and Sexual information} &
Sexual activity or curiosity; sexual or intimate body concerns; pregnancy;
fertility; reproductive status; intimate physical experiences; private bodily
functions. \\

7 &
\textbf{Romantic and intimate relationships} &
Spouses or partners; ex-partners; crushes; dating; attraction; cheating;
breakups; jealousy; relationship doubts; partner conflict. \\

8 &
\textbf{Family and close-relation information} &
Children; parents; siblings; relatives; dependents; household members; private
information about close friends. Romantic relationships are excluded. \\

9 &
\textbf{Risky interpersonal relationships} &
Conflict with friends, coworkers, classmates, neighbors, roommates, or
acquaintances; friendship problems; socially risky interactions. \\

10 &
\textbf{Emotionally vulnerable disclosure} &
Loneliness; fear; guilt; shame; despair; humiliation; emotional dependence;
feeling emotionally overwhelmed. \\

11 &
\textbf{Negative self-image} &
Feeling unattractive, incompetent, or unsuccessful; shame about one's body,
sexuality, work, or school performance; negative social comparison. \\

12 &
\textbf{Adverse life events and traumatic experiences} &
Abuse; harassment; victimization; bereavement; family crises; serious accidents;
violence; coercion; eviction; discrimination; death of family members or other
close relations. \\

13 &
\textbf{Stigmatized or taboo activities} &
Alcohol or drug use; intoxication; addiction or recovery; adult-content work;
taboo forms of work or activity; rule-breaking; other regulated or stigmatized
behavior. \\

14 &
\textbf{Cultural and ethnic background} &
Race or ethnicity; cultural identity; ancestry; migration or family background;
cultural stigma; racialized experiences; heritage-related family information. \\

15 &
\textbf{Ideological beliefs} &
Religious beliefs or upbringing; spirituality; political views; party
preference; activism; moral beliefs; controversial ideological positions. \\

\bottomrule
\end{tabularx}
\end{table*}

\section{Participants demographic}
\label{app:participant-demographics}

Table~\ref{tab:participant-demographics} summarizes the demographic
characteristics of the final survey sample and user-evaluation sample.

\begin{table}[t]
\centering
\caption{Demographics of Survey and User Evaluation Participants}
\label{tab:participant-demographics}
\footnotesize
\renewcommand{\arraystretch}{1.13}

\begin{minipage}[t]{0.47\columnwidth}
\centering
\textbf{Survey Participants (N = 146)}

\vspace{2pt}
\begin{tabular}{p{0.31\columnwidth} c}
\toprule
\textbf{Demographics} & \textbf{n (\%)} \\
\midrule
\textit{Age Range} & \\
18--24 & 22 (15.1\%) \\
25--34 & 61 (41.8\%) \\
35--44 & 28 (19.2\%) \\
45--54 & 23 (15.8\%) \\
55--64 & 7 (4.8\%) \\
65+ & 5 (3.4\%) \\
\midrule
\textit{Gender} & \\
Female & 79 (54.1\%) \\
Male & 66 (45.2\%) \\
Prefer not to say & 1 (0.7\%) \\
\midrule
\textit{Ethnicity} & \\
White & 103 (70.5\%) \\
Black & 19 (13.0\%) \\
Asian & 14 (9.6\%) \\
Mixed & 6 (4.1\%) \\
Other & 4 (2.7\%) \\
\bottomrule
\end{tabular}
\end{minipage}
\hfill
\begin{minipage}[t]{0.47\columnwidth}
\centering
\textbf{User Evaluation Participants (N = 15)}

\vspace{2pt}
\begin{tabular}{p{0.31\columnwidth} c}
\toprule
\textbf{Demographics} & \textbf{n (\%)} \\
\midrule
\textit{Age Range} & \\
18--24 & 2 (13.3\%) \\
25--34 & 6 (40.0\%) \\
35--44 & 6 (40.0\%) \\
45--54 & 1 (6.7\%) \\
\midrule
\textit{Gender} & \\
Female & 6 (40.0\%) \\
Male & 9 (60.0\%) \\
\midrule
\textit{Ethnicity} & \\
White & 7 (46.7\%) \\
Asian & 5 (33.3\%) \\
Black or African American & 4 (26.7\%) \\
American Indian or Alaska Native & 1 (6.7\%) \\
Hispanic or Latino/a/x & 1 (6.7\%) \\
Middle Eastern or North African & 1 (6.7\%) \\
\midrule
\textit{Education} & \\
High School Diploma or Equivalent & 2 (13.3\%) \\
Bachelor's Degree & 3 (20.0\%) \\
Master's Degree & 8 (53.3\%) \\
Doctoral or Professional Degree & 1 (6.7\%) \\
Another Educational Background & 1 (6.7\%) \\
\bottomrule
\end{tabular}
\end{minipage}

\vspace{3pt}
\raggedright
\scriptsize\textit{Note.} Participants could select multiple ethnic identities; therefore, ethnicity frequencies and percentages do not sum to the total sample size.
\end{table}

\section{User Evaluation Protocol}
\label{app:evaluation-protocol}

This appendix provides the moderator guide for the remote user
evaluation. Sessions lasted approximately 60 minutes and were conducted
over Zoom. Participants interacted with the prototype through remote
control of the researcher's shared screen. The moderator followed the
protocol below while allowing participants to ask questions, think
aloud, and decline to discuss any personal experience.

\subsection{Preparation}

Before each session, the researcher prepared two versions of the
prototype, each populated with the same fictional account history:
10--12 saved memory or profile items and 6--8 prior conversations.
This history situated the study task within an established account
rather than presenting the target disclosure as an isolated memory item.

The researcher also prepared a fictional task scenario based on the
participant's pre-study incident description. Scenarios preserved the
type of information and reason for wanting to forget it, while removing
identifying details and avoiding the participant's own sensitive
content. Each scenario included an initial prompt for the participant
to send to the assistant. We used AI to generate the prompt based on participants' description. 

\subsection{Introduction and Consent}

The moderator introduced the study as research on how people manage
what a conversational AI remembers. Participants were reminded that the
session would be recorded for research purposes, that only the research
team would access the recording, and that they could skip any question
or stop participation at any time. After confirming consent and
permission to record, the moderator began the warm-up interview.

\subsection{Warm-Up Interview}

The moderator asked participants the following questions:

\begin{itemize}
    \item What conversational AI systems have you used, and what do you
    usually use them for?
    \item What is your understanding of AI memory? How did you learn
    about it?
    \item You previously indicated that you had wanted an AI to forget
    something you shared. Could you describe one such experience?
    \item What were you trying to accomplish in that conversation?
    \item Why did you not want the AI to retain that information?
\end{itemize}

Participants were reminded that they did not need to disclose the
underlying sensitive information in detail.

\subsection{Comparative Evaluation}

Each participant used both conditions in counterbalanced order:
\textit{Baseline} (C0) and \textit{FENCE} (C1). The moderator
introduced the fictional account history and asked participants to
imagine that it belonged to them. Participants were encouraged to think
aloud and to interact as they would outside the study. There were no
right or wrong actions. The participant recieve the same task in both condition to ensure the study control. The example script apply for both C0 and C1 condition.

\subsubsection{Example Script}

Imagine that [Secanrio]you have been having some difficulties in a romantic relationship and are unsure how to approach a conversation with your partner. You decide to ask the AI for advice on how to communicate your concerns constructively.

You send this to AI [Prompt] ``I’ve been feeling hurt because my partner has repeatedly cancelled plans with me at the last minute, but I’m worried that bringing it up will sound accusatory. I still care about the relationship and want to understand what is going on rather than start an argument. Can you help me figure out how to talk to them about this and suggest what I could say?”

[Task] Imagine that this conversation contains information you would prefer
the AI not to store or use in future conversations. Please explore the system and take any action you think is appropriate. You may also choose not to take action. Let me know when you are finished.

\subsubsection{Post-Condition Questionnaire}
\label{app:evaluation-questionnaire}

After completing each system condition, participants completed the same
post-condition questionnaire, shown in Table \ref{tab:post-condition-questionnaire}, about the system they had just used. Unless
otherwise specified, items used a five-point Likert scale
(1 = \emph{Strongly disagree}; 5 = \emph{Strongly agree}). Items marked
(R) were reverse-coded for analysis.

\begin{table*}[t]
\centering
\caption{Post-condition questionnaire.}
\label{tab:post-condition-questionnaire}
\small
\renewcommand{\arraystretch}{1.15}
\setlength{\tabcolsep}{6pt}
\begin{tabularx}{\textwidth}{@{}p{0.20\textwidth}X@{}}
\toprule
\textbf{Measure} & \textbf{Item} \\
\midrule

Usability (UMUX; 7-point scale) &
This system's capabilities meet my requirements. \\

&
Using this system is a frustrating experience. (R) \\

&
This system is easy to use. \\

&
I have to spend too much time correcting things with this system. (R) \\
\midrule

Mental effort (9-point scale) &
How much mental effort did you have to put into managing what the system
should and should not save in memory? \\
\midrule

Memory interpretability (5-point scale) &
The system's memory state was always clear to me. \\

&
The system handled information in memory unpredictably. (R) \\

&
I was able to understand why information was or was not saved in memory. \\

&
It was difficult to identify what the system would do with information next. (R) \\
\midrule

Trust in memory management (5-point scale) &
I trust the system to manage memory as described. \\

&
I can rely on the system to carry out its stated memory-management behavior. \\
\bottomrule
\end{tabularx}
\end{table*}

For the usability items, participants responded on a seven-point agreement
scale (1 = \emph{Strongly disagree}; 7 = \emph{Strongly agree}). The mental
effort item used a nine-point scale ranging from 1 (\emph{Very, very low
mental effort}) to 9 (\emph{Very, very high mental effort}), with 5 labeled
\emph{Neither low nor high mental effort}.

\subsection{Follow-Up Exploration and Interview}

After completing both conditions and questionnaires, participants
returned to FENCE for a semi-structured interview and additional
exploration. The moderator first asked about actions taken during the
comparative tasks and probed any notable differences in questionnaire
responses. For example:

\begin{itemize}
    \item I noticed that you selected [category] / did not take action
    during the task. Could you explain your reasoning?
    \item You rated [construct] differently across the two systems.
    What led to that difference?
\end{itemize}

The moderator then invited participants to explore features they had
not used during the task. The researcher could provide messages from a
prompt bank spanning the 15 sensitive-information categories and
non-sensitive content when participants did not have an example in
mind. The moderator probed the following topics as relevant to the
participant's interaction:

\paragraph{Category-based protection.}
\begin{itemize}
    \item What is your understanding of what selecting categories does?
    \item Which categories would you choose to protect, and why?
    \item Did the categories cover the conversations you would not want
    an AI to remember? What, if anything, was missing or unclear?
    \item How would category-based protection help or fail to help in
    situations like the one you discussed earlier?
\end{itemize}

\paragraph{Vanishing and protected chats.}
\begin{itemize}
    \item What is your understanding of vanishing chat and protected
    chat?
    \item In what situations would you use each mode?
    \item How did you understand the relationship between these modes,
    visible chat history, and AI memory?
    \item Was the current protection status easy to notice and
    understand? Why or why not?
\end{itemize}

\paragraph{Comparison with existing practices.}
\begin{itemize}
    \item How does this system compare with how you would currently
    manage AI memory?
    \item Did you feel that your information was protected? Why or why
    not?
    \item Would you use a system like this if it were available? Why?
    \item What would you change about the system?
\end{itemize}

The session concluded with an invitation for additional comments and a
demographic questionnaire.

\section{Item-Level Quantitative Results}
\label{app:item-level-stats}

Table~\ref{tab:ratings-items} reports item-level Wilcoxon signed-rank
statistics for each measure in the post-condition questionnaire
(Appendix~\ref{app:evaluation-questionnaire}), comparing the baseline
condition (C0) and the gated-memory condition (C1). Composite-level
results are reported in Table~\ref{tab:ratings} in
Section~\ref{sec:quantitative-result}.

\begin{table}[t]
\centering
\caption{Item-level ratings comparing the baseline condition (C0) and
the gated-memory condition (C1). Wilcoxon signed-rank tests,
Bonferroni-corrected across the four overall measures; $N=15$. Items
marked (R) were reverse-coded before analysis.}
\label{tab:ratings-items}
\begin{tabular}{llccccc}
\toprule
\textbf{Measure} & \textbf{Item} & \textbf{Mdn (C0)} & \textbf{Mdn (C1)} & \textbf{$Z$} & \textbf{$p_{\mathrm{adj}}$} & \textbf{$r$} \\
\midrule
Usability & Capabilities meet requirements & 5.00 & 6.00 & $-2.57$ & $.041$ & .74 \\
Usability & Frustrating experience (R)     & 6.00 & 7.00 & $-2.32$ & $.081$ & .77 \\
Usability & Easy to use                    & 6.00 & 6.00 & $-0.96$ & $1.00$ & .39 \\
Usability & Time correcting (R)            & 6.00 & 6.00 & $-1.19$ & $.937$ & .45 \\
\midrule
Trust & Manage memory as described  & 3.00 & 4.00 & $-2.53$ & $.046$ & .76 \\
Trust & Rely on stated behavior     & 3.00 & 4.00 & $-2.65$ & $.033$ & .76 \\
\midrule
Interpretability & Memory state clear                & 2.00 & 4.00 & $-2.24$ & $.100$ & .62 \\
Interpretability & Handled unpredictably (R)          & 4.00 & 5.00 & $-2.12$ & $.135$ & .75 \\
Interpretability & Understood why saved/not saved     & 4.00 & 5.00 & $-2.75$ & $.024$ & .92 \\
Interpretability & Difficult to identify next (R)     & 3.00 & 5.00 & $-1.79$ & $.297$ & .54 \\
\midrule
Mental Effort & (single item) & 4.00 & 3.00 & $-1.80$ & $.288$ & .60 \\
\bottomrule
\end{tabular}
\end{table}

\section{User Evaluation Codebook}
\label{app:qualitative-codebook}

Table~\ref{tab:qualitative-codebook} presents the codebook used in the
thematic analysis of participants' think-aloud comments, interview responses,
and observation notes from the study sessions. Codes were not mutually
exclusive; a single excerpt could receive multiple codes.
{
\scriptsize
\setlength{\tabcolsep}{3pt}
\renewcommand{\arraystretch}{0.92}
\setlength{\LTpre}{2pt}
\setlength{\LTpost}{2pt}

\begin{longtable}{@{}p{0.19\textwidth}p{0.29\textwidth}p{0.46\textwidth}@{}}
\caption{Qualitative codebook for the user evaluation.}
\label{tab:qualitative-codebook}\\
\toprule
\textbf{Theme} & \textbf{Code} & \textbf{Description} \\
\midrule
\endfirsthead

\toprule
\textbf{Theme} & \textbf{Code} & \textbf{Description} \\
\midrule
\endhead

\bottomrule
\endfoot

\multicolumn{3}{@{}l}{\textit{Design tradeoffs and protection mechanisms}} \\
\midrule
&
Balance between under-protection and over-protection &
Comments about the tradeoff between protecting too much information and
failing to protect information that users consider sensitive, including
expectations about whether AI should remember by default or only when
explicitly instructed. \\

&
Malleable data unit &
Comments about the appropriate unit of protection or memory management, such
as an entire conversation, individual message, sentence, phrase, or a specific
future use of information. \\

&
Control granularity/system automation &
Comments about the desired balance between user-configurable control and
automated protection, including when users want finer-grained choices or
prefer the system to act on their behalf. \\

&
Different ways of protection &
Participant proposals for alternative protection mechanisms, such as
redaction, masking, deletion, restricted reuse, or protection that does not
require removing a conversation. \\

&
Default mechanism &
Expectations or preferences regarding what should happen by default, including
whether conversations should enter memory unless users opt out or remain
excluded unless users opt in. \\

&
Flexibility in protection &
General comments about the system's ability to accommodate different
protection needs through protected chats, vanishing chats, or cancellation. \\

&
Feature: protected chat &
Comments about the purpose, value, or limitations of keeping a conversation
visible in history while excluding it from AI memory. \\

&
Feature: vanishing chat &
Comments about the purpose, value, or limitations of automatically removing a
protected conversation after a selected retention period. \\

\midrule
\multicolumn{3}{@{}l}{\textit{Current memory practices and mental models}} \\
\midrule
&
Experience with memory &
Participants' positive or negative prior experiences with conversational AI
memory and personalization, including experiences not directly related to the
target scenario. \\

&
Mental model of memory &
Participants' general understanding or assumptions about how commercial AI
systems retain, retrieve, and use information across conversations. \\

&
Management: prompting AI to forget as a conversational workaround &
Expectations, reasons, and prior experiences of directly asking an AI to
forget information, including why this conversational action felt more
available than other controls. \\

&
Management: deletion does not guarantee forgetting &
Participants' distinction between deleting a chat or memory item and ensuring
that the AI no longer retains or uses the information. \\

&
Management: deletion can delete too much &
Concerns that deleting a conversation may also remove useful context,
information, or records that participants want to keep. \\

&
Management: undiscoverability of current ways of management &
Reasons participants did not know about, notice, or consider existing
post-sharing memory controls, particularly during the baseline condition. \\

&
Management: helplessness due to the black box &
Feelings that AI memory is opaque or uncontrollable, such that participants
could not verify or meaningfully influence what the system retained. \\

&
Management: self-censorship &
Prior decisions to withhold, alter, or avoid sharing information because of
concerns about AI retention, including occasions when self-censorship failed. \\

&
Mental model of memory: concerns &
Specific concerns that follow from participants' understanding of AI memory,
such as unwanted reuse, inference, interpersonal exposure, or future effects
on system responses. \\

\midrule
\multicolumn{3}{@{}l}{\textit{Perceived value of gated memory}} \\
\midrule
&
Useful: scenario &
Specific situations or use cases in which participants considered the system
helpful. \\

&
Useful: transparency &
Comments that the system made memory behavior, protection status, or the
reason for a decision easier to understand. \\

&
Useful: feeling in control &
Comments that the system gave participants greater perceived authority over
what the AI could retain or use later. \\

&
Useful: awareness &
Comments that category selection or system feedback increased participants'
awareness of sensitive disclosures or encouraged reflection before sharing. \\

&
Useful: saving effort &
Comments that the system could reduce the effort of managing memory, either
during the task or over continued use. \\

&
Useful: assurance and security &
Comments that the system provided reassurance, peace of mind, or a greater
sense of security about how sensitive information would be handled. \\

&
Useful: policy-level protection &
Comments that the category-based gate provided a standing policy that the
system could follow, rather than requiring participants to intervene after
each disclosure. \\

&
Useful: willingness to adopt in the future &
Explicit statements that participants would adopt, continue using, or
recommend the system in future use. \\

\midrule
\multicolumn{3}{@{}l}{\textit{Trust in system behavior}} \\
\midrule
&
Trust &
General perceptions of whether the system can be trusted to handle
conversations and memory according to its stated behavior. This parent code
was used when a comment did not concern a more specific source of trust below. \\

&
Trust: protection before memory entry &
Trust arising from the system preventing selected sensitive information from
entering memory in the first place, rather than requiring users to discover,
remove, or remediate retained information afterward. \\

&
Trust: accurate result &
Trust arising from the system detecting categories and applying protection in
ways that matched participants' expectations. \\

&
Trust: robust system &
Trust contingent on the system consistently detecting sensitive information
accurately and reliably across conversations and contexts. \\

&
Trust: auditability/ability to verify &
Desire to verify that protection worked, for example by inspecting chat
history, memory records, system feedback, or behavior in later conversations. \\

&
Trust: provider versus local/client-side boundary &
Comments about how the location of processing, the service provider, or
client-side operation shaped trust in the system. \\

&
Trust: transparency &
Trust arising from the system making its memory behavior visible or
understandable, such as showing why protection was triggered or whether a
conversation was excluded from memory. \\

\midrule
\multicolumn{3}{@{}l}{\textit{Selecting categories to protect}} \\
\midrule
&
Learning curve &
Experiences of added effort, confusion, or familiarization required to learn
the category-based controls or the system's protection modes. \\

&
What to protect: analogy to AI as a human &
Reasoning about category selection by imagining the AI as a person, such as a
friend, stranger, colleague, or someone who should not remember the
information. \\

&
What to protect: one-time use &
Reasoning about whether information would be useful only for the current task
or should remain available for future conversations. \\

&
What to protect: no anticipated need to share &
Decisions not to select a category because participants did not expect to
share that type of information, often because they would avoid disclosing it
to an AI altogether. \\

&
What to protect: leave it protected anyway &
Selecting a category as a precaution even when participants did not personally
regard every disclosure in that category as sensitive. \\

&
What to protect: useful to remember despite sensitivity &
Reasoning that some sensitive information should remain memory-eligible
because it could improve future assistance, including the participant's final
selection decision. \\

&
What to protect: role/context separation &
Category-selection decisions that depended on the social role, task, audience,
or context in which the information was disclosed. \\

&
What to protect: potential harm from remembering &
Reasoning that information should be protected because its retention, reuse,
exposure, inference, or influence on future AI responses could cause harm. \\

&
Reasons for selecting or not selecting a category &
Explanations for choosing or declining a particular category that do not fit
the more specific selection rationales above. \\

&
Category selection creates privacy concern &
Concerns that the act of selecting protected categories could itself reveal
sensitive information or create a new privacy risk. \\

&
Taxonomy: balance of category granularity &
Comments about whether the taxonomy was too broad, too narrow, or otherwise
unable to represent participants' intended protection boundaries. \\

&
Taxonomy: added effort but necessary &
Acknowledgment that selecting categories increased effort, alongside a view
that this effort was necessary or worthwhile for meaningful control. \\

&
Taxonomy: positive category feedback &
Positive feedback about the availability, wording, organization, or interface
presentation of the categories. \\

\midrule
\multicolumn{3}{@{}l}{\textit{Explanations of post-condition ratings}} \\
\midrule
&
Survey: usability &
Explanations of usability ratings that reveal a notable experience or help
interpret an observed quantitative result. \\

&
Survey: interpretability &
Explanations of interpretability ratings that reveal a notable experience or
help interpret an observed quantitative result. \\

&
Survey: trust &
Explanations of trust ratings that reveal a notable experience or help
interpret an observed quantitative result. \\

&
Survey: mental effort &
Explanations of mental-effort ratings that reveal a notable experience or
help interpret an observed quantitative result. \\
\end{longtable}
}

\section{Model Classification Prompt}
\label{app:model-classification-prompt}

The following prompt was provided to each model without
task-specific training or fine-tuning. For every benchmark message,
the \texttt{\{\{MESSAGE\}\}} placeholder was replaced with the message
being evaluated. The model was instructed to return one binary
prediction for each of the 15 categories.

\begin{lstlisting}[
    style=classificationprompt,
    caption={Prompt used to obtain the 15 independent binary category predictions for each message.},
    label={lst:classification-prompt}
]
You are a classifier for a privacy-sensitive information taxonomy.
Classify the MESSAGE into 15 independent binary categories.
For each category:

- `1` = the message contains or directly concerns information covered by the category.
- `0` = the category is not sufficiently supported.

Multiple categories may be `1`.

## Decision Rules

1. Carefully read each category's definition and illustrative examples, and use both to accurately understand its intended scope and boundaries before assigning any labels.
2. A category can apply to a disclosure or to a request whose meaning directly depends on the sensitive circumstance or activity. Questions, help requests, writing requests, and hypothetical or fictional scenarios are not automatically negative.
3. Read the message carefully and understand its semantic meaning, not just match the keyword with illustrative examples or definitions.
4. Do not infer identities, conditions, beliefs, relationships, or circumstances that are not stated or directly supported.
5. Evaluate each category independently. The same information may support multiple categories.

## Categories

**C01 — Sexual, reproductive, and intimate bodily information**
Definition: Messages that disclose a person's sexual, reproductive and inimate bodily information or experiences.
Illustrative examples: Sexual activity or curiosity; sexual or intimate body concerns; pregnancy; fertility; reproductive status; intimate physical experiences; private bodily functions

**C02 — Risky interpersonal relationship and conflict**
Definition: Messages that disclose a person's sensitive dynamics, or tensions within non-romantic and not-within-family social relationships.
Illustrative examples: Conflict *with people who is not a family-member and not a romantic partner, including* friends, coworkers, classmates, neighbors, roommates, or acquaintances; social dynamic concerns or problems; socially risky interactions

**C03 — Emotionally vulnerable disclosure**
Definition: Messages that reveal a person's vulnerable emotional states or experiences.
Illustrative examples: Loneliness; fear; guilt; shame; despair; humiliation; emotional dependence; feeling emotionally overwhelmed, *not including emotions that are not vulnerable or descriptions that does not indicate a emotional status*

**C04 — Romantic and Intimate Relationships**
Definition: Messages that disclose sensitive information about a person's romantic, intimate, or desired relationships and their dynamics.
Illustrative examples: Spouses or partners; ex-partners; crushes; dating; attraction; cheating; breakup; jealousy; relationship doubts; partner conflict

**C05 — Legal and Administrative Proceedings**
Definition: Messages that disclose a person's involvement in the legal, governmental, administrative processes.
Illustrative examples: Court or police matters; tax or government processes; immigration; institutional complaints; disciplinary procedures; lawsuits; custody or divorce proceedings; formal investigations

**C06 — Personal identifying, location, account, and profiling information**
Definition: Messages that disclose information capable of identifying, locating, or contacting an individual.
Illustrative examples: Names; phone numbers; emails; addresses; location details; usernames; account information; school or workplace names; strong demographic signals; IP/network clues; excludes fictional character names and celebrity names

**C07 — Professional Identity and Career-Sensitive Information**
Definition: Messages that disclose specific professional or academic identity information, or information that could affect a person's career opportunities, standing, or reputation.
Illustrative examples: Employer or manager information; job search; performance issues; workplace conflict; professional reputation; confidential work context; career or academic history

**C08 — Family and Close-Relation Information**
Definition: Messages that disclose private information concerning a person's family members, dependents, household members, or other close relations.
Illustrative examples: Children; parents; siblings; relatives; dependents; household members; private information about close friends; *excluding information about romantic relationship, contents about husband and wife will not be included*

**C09 — Health information**
Definition: Messages that disclose information concerning a person's physical or explicit mental health disorder.
Illustrative examples: Symptoms; diagnoses; medications; therapy; disability; mental-health distress as an disorder, excluding normal stress or anxious feeling; medical records; treatment; insurance or medical billing; health or mental-health issue-related care-seeking, excluding normal desire for more emotional care or attention

**C10 — Financial information**
Definition: Messages that disclose information about a person's finances or financial activities.
Illustrative examples: Income; debt; loans; banking; credit cards; taxes; purchases; rent; benefits; financial hardship; family finances; financial conflict

**C11 — Adverse life events and traumatic experiences**
Definition: Messages that disclose experiences involving substantial harm, victimization, death, or other severe life events.
Illustrative examples: Abuse; harassment; victimization; bereavement; family crisis; serious accidents; violence; coercion; eviction; discrimination; death of family members or close-relation

**C12 — Negative self-image**
Definition: Messages that express negative evaluations of oneself or perceived personal inadequacy.
Illustrative examples: Feeling unattractive; incompetent; unsuccessful; ashamed of one's body, sexuality, work, or school performance; social comparison

**C13 — Cultural and Ethnic Background**
Definition: Messages that disclose a person's identity, background, or lived experience associated with race, ethnicity, culture, ancestry, or heritage.
Illustrative examples: Race or ethnicity; cultural identity; ancestry; migration or family background; cultural stigma; racialized experiences; heritage-related family information

**C14 — Ideological beliefs**
Definition: Messages that disclose someone's personal beliefs or commitments related to religion, politics or other ideology.
Illustrative examples: Religious beliefs or upbringing; spirituality; political views; party preference; activism; moral beliefs; controversial ideological positions

**C15 — Stigmatized or taboo activities**
Definition: Messages that disclose activities or behaviors that may carry substantial social stigma, moral judgment, or cultural taboos.
Illustrative examples: Alcohol or drug use; intoxication; addiction or recovery; adult-content work; taboo forms of work or activity; rule-breaking; other regulated or stigmatized behavior

## Output

Return ONLY the following JSON object with all 15 labels. Do not add explanations, Markdown, or additional text.
{
"C01": 0,
"C02": 0,
"C03": 0,
"C04": 0,
"C05": 0,
"C06": 0,
"C07": 0,
"C08": 0,
"C09": 0,
"C10": 0,
"C11": 0,
"C12": 0,
"C13": 0,
"C14": 0,
"C15": 0
}
MESSAGE:
{{MESSAGE}}
\end{lstlisting}

%% file: bibliography.bib
@inproceedings{Yang2024FutureHCIPolicy,
  author    = {Yang, Qian and Wong, Richmond Y. and Jackson, Steven J. and Junginger, Sabine and Hagan, Margaret D. and Gilbert, Thomas K. and Zimmerman, John},
  title     = {The Future of HCI-Policy Collaboration},
  booktitle = {Proceedings of the 2024 CHI Conference on Human Factors in Computing Systems},
  year      = {2024},
  pages     = {1--15},
  articleno = {820},
  publisher = {Association for Computing Machinery},
  doi       = {10.1145/3613904.3642771}
}

@misc{GDPR2016,
  author       = {{European Parliament and Council of the European Union}},
  title        = {Regulation (EU) 2016/679 of the European Parliament and of the Council of 27 April 2016 on the Protection of Natural Persons with Regard to the Processing of Personal Data and on the Free Movement of Such Data (General Data Protection Regulation)},
  year         = {2016},
  note         = {Article 9: Processing of special categories of personal data},
  howpublished = {Official Journal of the European Union, L 119}
}

@misc{EUAIAct2024,
  author       = {{European Parliament and Council of the European Union}},
  title        = {Regulation (EU) 2024/1689 Laying Down Harmonised Rules on Artificial Intelligence (Artificial Intelligence Act)},
  year         = {2024},
  howpublished = {Official Journal of the European Union}
}

@article{Zhang2025MemorySurvey,
  author    = {Zeyu Zhang and Quanyu Dai and Xiaohe Bo and Chen Ma and Rui Li and Xu Chen and Jieming Zhu and Zhenhua Dong and Ji-Rong Wen},
  title     = {A Survey on the Memory Mechanism of Large Language Model-based Agents},
  journal   = {ACM Transactions on Information Systems},
  volume    = {43},
  number    = {6},
  articleno = {155},
  pages     = {1--47},
  year      = {2025},
  publisher = {Association for Computing Machinery},
  doi       = {10.1145/3748302}
}

@inproceedings{Jones2025AgentMemory,
  author    = {Brennan Jones and Kelsey Stemmler and Emily Su and Young-Ho Kim and Anastasia Kuzminykh},
  title     = {Users' Expectations and Practices with Agent Memory},
  booktitle = {Extended Abstracts of the CHI Conference on Human Factors in Computing Systems},
  series    = {CHI EA '25},
  articleno = {563},
  numpages  = {8},
  year      = {2025},
  publisher = {Association for Computing Machinery},
  doi       = {10.1145/3706599.3720158}
}

@inproceedings{Huang2023MemorySandbox,
  author    = {Ziheng Huang and Sebastian Gutierrez and Hemanth Kamana and Stephen MacNeil},
  title     = {Memory Sandbox: Transparent and Interactive Memory Management for Conversational Agents},
  booktitle = {Adjunct Proceedings of the 36th Annual ACM Symposium on User Interface Software and Technology},
  series    = {UIST '23 Adjunct},
  articleno = {97},
  numpages  = {3},
  year      = {2023},
  publisher = {Association for Computing Machinery},
  doi       = {10.1145/3586182.3615796}
}

@inproceedings{Kim2026MetaphorsMemory,
  author    = {Munyeong Kim and Michalis Famelis and Ian Arawjo},
  title     = {Metaphors for Memory: Charting a Design Space of AI Memory Tools and Interfaces},
  booktitle = {Proceedings of the 2026 ACM Designing Interactive Systems Conference},
  series    = {DIS '26},
  pages     = {2955--2971},
  year      = {2026},
  publisher = {Association for Computing Machinery},
  doi       = {10.1145/3800645.3812979}
}

@article{Solove2006Taxonomy,
  author  = {Solove, Daniel J.},
  title   = {A Taxonomy of Privacy},
  journal = {University of Pennsylvania Law Review},
  year    = {2006},
  volume  = {154},
  number  = {3},
  pages   = {477--564},
  doi     = {10.2307/40041279}
}

@article{Nissenbaum2004Privacy,
  author  = {Nissenbaum, Helen},
  title   = {Privacy as Contextual Integrity},
  journal = {Washington Law Review},
  year    = {2004},
  volume  = {79},
  number  = {1},
  pages   = {119--158},
  url     = {https://digitalcommons.law.uw.edu/wlr/vol79/iss1/10/}
}

@article{Nissenbaum2019ContextualIntegrity,
  author  = {Nissenbaum, Helen},
  title   = {Contextual Integrity Up and Down the Data Food Chain},
  journal = {Theoretical Inquiries in Law},
  year    = {2019},
  volume  = {20},
  number  = {1},
  pages   = {221--256},
  doi     = {10.1515/til-2019-0008}
}

@article{Thompson2005Sensitive,
  author  = {Thompson, E. Dale and Kaarst-Brown, Michelle L.},
  title   = {Sensitive Information: A Review and Research Agenda},
  journal = {Journal of the American Society for Information Science and Technology},
  year    = {2005},
  volume  = {56},
  number  = {3},
  pages   = {245--257},
  doi     = {10.1002/asi.20121}
}

@article{Ohm2015Sensitive,
  author  = {Ohm, Paul},
  title   = {Sensitive Information},
  journal = {Southern California Law Review},
  year    = {2015},
  volume  = {88},
  number  = {5},
  pages   = {1125--1196}
}

@article{Milne2017InformationSensitivity,
  author  = {Milne, George R. and Pettinico, George and Hajjat, Fatima M. and Markos, Ereni},
  title   = {Information Sensitivity Typology: Mapping the Degree and Type of Risk Consumers Perceive in Personal Data Sharing},
  journal = {Journal of Consumer Affairs},
  year    = {2017},
  volume  = {51},
  number  = {1},
  pages   = {133--161},
  doi     = {10.1111/joca.12111}
}

@article{Saglam2022PersonalInformation,
  author  = {Sa{\u g}lam, Rahime Belen and Nurse, Jason R. C. and Hodges, Duncan},
  title   = {Personal Information: Perceptions, Types and Evolution},
  journal = {Journal of Information Security and Applications},
  year    = {2022},
  volume  = {66},
  pages   = {103163},
  doi     = {10.1016/j.jisa.2022.103163}
}

@misc{Zaman2026InferentialPrivacy,
  author        = {Zaman, S M Mehedi and Garimella, Kiran},
  title         = {Inferential Privacy Leakage in Anonymized Conversational AI Logs},
  year          = {2026},
  eprint        = {2605.23820},
  archivePrefix = {arXiv},
  primaryClass  = {cs.CY},
  url           = {https://arxiv.org/abs/2605.23820}
}

@techreport{NISTSP80060,
  author      = {Stine, Kevin and Kissel, Richard and Barker, William C. and Fahlsing, Jim and Gulick, Jessica},
  title       = {Guide for Mapping Types of Information and Information Systems to Security Categories},
  institution = {National Institute of Standards and Technology},
  number      = {NIST Special Publication 800-60 Volume I Revision 1},
  year        = {2008},
  doi         = {10.6028/NIST.SP.800-60v1r1}
}

@techreport{chatterji2025people,
  title={How people use chatgpt},
  author={Chatterji, Aaron and Cunningham, Thomas and Deming, David J and Hitzig, Zoe and Ong, Christopher and Shan, Carl Yan and Wadman, Kevin},
  year={2025},
  institution={National Bureau of Economic Research}
}

@inproceedings{Ngong2025ContextualPrivacy,
  author    = {Ngong, Ivoline C. and Kadhe, Swanand Ravindra and Wang, Hao and Murugesan, Keerthiram and Weisz, Justin D. and Dhurandhar, Amit and Natesan Ramamurthy, Karthikeyan},
  title     = {Protecting Users From Themselves: Safeguarding Contextual Privacy in Interactions with Conversational Agents},
  booktitle = {Findings of the Association for Computational Linguistics: ACL 2025},
  year      = {2025},
  pages     = {26196--26220},
  publisher = {Association for Computational Linguistics},
  address   = {Vienna, Austria},
  month     = jul,
  doi       = {10.18653/v1/2025.findings-acl.1343},
  url       = {https://aclanthology.org/2025.findings-acl.1343/}
}

@inproceedings{Li2026PrivacyControl,
  author    = {Li, Zhuoyang and Wu, Yanlai and Li, Yao and Gui, Xinning and Luo, Yuhan},
  title     = {Privacy Control in Conversational LLM Platforms: A Walkthrough Study},
  booktitle = {Proceedings of the 2026 CHI Conference on Human Factors in Computing Systems},
  year      = {2026},
  publisher = {Association for Computing Machinery},
  address   = {New York, NY, USA},
  articleno = {535},
  numpages  = {26},
  doi       = {10.1145/3772318.3791054},
  url       = {https://doi.org/10.1145/3772318.3791054}
}

@article{Packer2023MemGPT,
  author  = {Packer, Charles and Wooders, Sarah and Lin, Kevin and Fang, Vivian and Patil, Shishir G. and Stoica, Ion and Gonzalez, Joseph E.},
  title   = {{MemGPT}: Towards LLMs as Operating Systems},
  journal = {arXiv preprint arXiv:2310.08560},
  year    = {2023}
}

@inproceedings{Zhong2024MemoryBank,
  author    = {Zhong, Wanjun and Guo, Lianghong and Gao, Qiqi and Ye, He and Wang, Yanlin},
  title     = {MemoryBank: Enhancing Large Language Models with Long-Term Memory},
  booktitle = {Proceedings of the AAAI Conference on Artificial Intelligence},
  volume    = {38},
  number    = {17},
  pages     = {19724--19731},
  year      = {2024},
  doi       = {10.1609/AAAI.V38I17.29946}
}

@misc{CCPA,
  author = {{State of California}},
  title  = {California Consumer Privacy Act of 2018, as Amended},
  year   = {2018},
  note   = {California Civil Code Sections 1798.100--1798.199.100}
}

@inproceedings{Dou2024SelfDisclosure,
  author    = {Dou, Yao and Krsek, Isadora and Naous, Tarek and Kabra, Anubha and Das, Sauvik and Ritter, Alan and Xu, Wei},
  title     = {Reducing Privacy Risks in Online Self-Disclosures with Language Models},
  booktitle = {Proceedings of the 62nd Annual Meeting of the Association for Computational Linguistics (Volume 1: Long Papers)},
  year      = {2024},
  pages     = {13732--13754},
  publisher = {Association for Computational Linguistics},
  address   = {Bangkok, Thailand},
  doi       = {10.18653/v1/2024.acl-long.741}
}

@inproceedings{Park2023GenerativeAgents,
  author    = {Park, Joon Sung and O'Brien, Joseph C. and Cai, Carrie J. and Morris, Meredith Ringel and Liang, Percy and Bernstein, Michael S.},
  title     = {Generative Agents: Interactive Simulacra of Human Behavior},
  booktitle = {Proceedings of the 36th Annual ACM Symposium on User Interface Software and Technology},
  year      = {2023},
  articleno = {2},
  pages     = {1--22},
  publisher = {Association for Computing Machinery},
  doi       = {10.1145/3586183.3606763}
}

@article{Lucas2014Disclosure,
  author  = {Lucas, Gale M. and Gratch, Jonathan and King, Aisha and Morency, Louis-Philippe},
  title   = {It's Only a Computer: Virtual Humans Increase Willingness to Disclose},
  journal = {Computers in Human Behavior},
  volume  = {37},
  pages   = {94--100},
  year    = {2014},
  doi     = {10.1016/j.chb.2014.04.043}
}

@article{Ho2018ChatbotDisclosure,
  author  = {Ho, Annabell and Hancock, Jeff and Miner, Adam S.},
  title   = {Psychological, Relational, and Emotional Effects of Self-Disclosure After Conversations With a Chatbot},
  journal = {Journal of Communication},
  volume  = {68},
  number  = {4},
  pages   = {712--733},
  year    = {2018},
  doi     = {10.1093/joc/jqy026}
}

@inproceedings{Staab2024BeyondMemorization,
  author    = {Staab, Robin and Vero, Mark and Balunovi{\'c}, Mislav and Vechev, Martin},
  title     = {Beyond Memorization: Violating Privacy via Inference with Large Language Models},
  booktitle = {The Twelfth International Conference on Learning Representations},
  year      = {2024}
}

@inproceedings{Zhang2025RAGMemoryPrivacy,
  author    = {Zhang, Shuning and Ma, Rongjun and Ma, Ying and Li, Shixuan and Xu, Yiqun and Yi, Xin and Li, Hewu},
  title     = {Understanding Users' Privacy Perceptions Towards LLM's RAG-based Memory},
  booktitle = {Proceedings of the 2025 Workshop on Human-Centered AI Privacy and Security},
  pages     = {10--19},
  year      = {2025},
  publisher = {Association for Computing Machinery},
  doi       = {10.1145/3733816.3760750}
}

@inproceedings{Yen2024Memolet,
  author    = {Yen, Ryan and Zhao, Jian},
  title     = {Memolet: Reifying the Reuse of User-AI Conversational Memories},
  booktitle = {Proceedings of the 37th Annual ACM Symposium on User Interface Software and Technology},
  year      = {2024},
  articleno = {58},
  pages     = {1--22},
  publisher = {Association for Computing Machinery},
  doi       = {10.1145/3654777.3676388}
}

@inproceedings{Vaithilingam2025SemanticCommit,
  author    = {Vaithilingam, Priyan and Kim, Munyeong and Acosta-Parenteau, Frida-Cecilia and Lee, Daniel and Mhedhbi, Amine and Glassman, Elena L. and Arawjo, Ian},
  title     = {Semantic Commit: Helping Users Update Intent Specifications for AI Memory at Scale},
  booktitle = {Proceedings of the 38th Annual ACM Symposium on User Interface Software and Technology},
  year      = {2025},
  articleno = {137},
  pages     = {1--18},
  publisher = {Association for Computing Machinery},
  doi       = {10.1145/3746059.3747778}
}

@misc{OpenAI2026MemoryFAQ,
  author       = {{OpenAI}},
  title        = {Memory FAQ},
  year         = {2026},
  howpublished = {OpenAI Help Center},
  url          = {https://help.openai.com/en/articles/8590148},
  note         = {Accessed September 6, 2026}
}

@misc{Google2026GeminiMemory,
  author       = {{Google}},
  title        = {Get Personalization with Memory of Your Past Gemini Chats},
  year         = {2026},
  howpublished = {Gemini Apps Help},
  url          = {https://support.google.com/gemini/answer/16598469},
  note         = {Accessed September 6, 2026}
}

@misc{Anthropic2026ClaudeMemory,
  author       = {{Anthropic}},
  title        = {Use Claude's Chat Search and Memory to Build on Previous Context},
  year         = {2026},
  howpublished = {Claude Help Center},
  url          = {https://support.claude.com/en/articles/11817273-use-claude-s-chat-search-and-memory-to-build-on-previous-context},
  note         = {Accessed September 6, 2026}
}

@misc{Google2026TemporaryChat,
  author       = {{Google}},
  title        = {Use Gemini Apps},
  year         = {2026},
  howpublished = {Gemini Apps Help},
  url          = {https://support.google.com/gemini/answer/13275745},
  note         = {Section: Start a temporary chat; Accessed September 6, 2026}
}

@misc{NISTAgentStandards2026,
  author = {{National Institute of Standards and Technology}},
  title = {Announcing the {AI} Agent Standards Initiative for Interoperable and Secure Innovation},
  year = {2026},
  month = feb,
  url = {https://www.nist.gov/news-events/news/2026/02/announcing-ai-agent-standards-initiative-interoperable-and-secure}
}

@inproceedings{South2025Delegation,
  author = {South, Tobin and Marro, Samuele and Hardjono, Thomas and Mahari, Robert and Whitney, Cedric Deslandes and Chan, Alan and Pentland, Alex},
  title = {Position: {AI} Agents Need Authenticated Delegation},
  booktitle = {Proceedings of the 42nd International Conference on Machine Learning},
  series = {Proceedings of Machine Learning Research},
  volume = {267},
  pages = {82211--82231},
  year = {2025},
  publisher = {PMLR},
  url = {https://proceedings.mlr.press/v267/south25a.html}
}

@misc{OWASPAgentic2025,
  author = {{OWASP GenAI Security Project}},
  title = {{OWASP Top 10 for Agentic Applications for 2026}},
  year = {2025},
  month = dec,
  url = {https://genai.owasp.org/resource/owasp-top-10-for-agentic-applications-for-2026/},
  note = {2026 edition, released December 2025}
}

@techreport{NISTAIRMF2023,
  author = {Tabassi, Elham},
  title = {Artificial Intelligence Risk Management Framework ({AI RMF 1.0})},
  institution = {National Institute of Standards and Technology},
  number = {NIST AI 100-1},
  year = {2023},
  doi = {10.6028/NIST.AI.100-1},
  url = {https://doi.org/10.6028/NIST.AI.100-1}
}

@inproceedings{Wang2025MemoryPrivacy,
  author = {Wang, Bo and He, Weiyi and Zeng, Shenglai and Xiang, Zhen and Xing, Yue and Tang, Jiliang and He, Pengfei},
  title = {Unveiling Privacy Risks in {LLM} Agent Memory},
  booktitle = {Proceedings of the 63rd Annual Meeting of the Association for Computational Linguistics (Volume 1: Long Papers)},
  pages = {25241--25260},
  year = {2025},
  address = {Vienna, Austria},
  publisher = {Association for Computational Linguistics},
  doi = {10.18653/v1/2025.acl-long.1227},
  url = {https://aclanthology.org/2025.acl-long.1227/}
}

@misc{Microsoft2026Memory,
  author = {Isak, Natalie and Cooley, Sarah},
  title = {Guarding {AI} Memory},
  year = {2026},
  month = jun,
  howpublished = {Microsoft Security Blog},
  url = {https://www.microsoft.com/en-us/security/blog/2026/06/22/guarding-ai-memory/}
}

@misc{Pulipaka2026Poisoning,
  author = {Pulipaka, Sidharth and Hlebik, Stanislau and Raghav, Leonidas and Abdelnabi, Sahar and Raina, Vyas and Sheth, Ivaxi and Fritz, Mario},
  title = {Hidden in Memory: Sleeper Memory Poisoning in {LLM} Agents},
  year = {2026},
  eprint = {2605.15338},
  archivePrefix = {arXiv},
  primaryClass = {cs.CR},
  url = {https://arxiv.org/abs/2605.15338}
}

@misc{Meta2026Muse,
  author = {{Meta}},
  title = {Introducing Muse: The World's First Personal {AI} Agent Built for Everyone},
  year = {2026},
  month = sep,
  howpublished = {Meta Newsroom},
  url = {https://about.fb.com/news/2026/09/introducing-muse-personal-ai-agent/}
}

@misc{Meta2026MuseDesign,
  author = {Sarantakos, Mona and Awad, Christine},
  title = {How We Designed Muse},
  year = {2026},
  month = sep,
  howpublished = {Meta},
  url = {https://introducing.muse.ai/}
}

@misc{Mitchell2026Oversight,
  author = {Mitchell, Margaret and Ghosh, Avijit and Passi, Samir},
  title = {{AI} Agents Push Humans Out of the Loop},
  year = {2026},
  eprint = {2608.23642},
  archivePrefix = {arXiv},
  url = {https://arxiv.org/abs/2608.23642}
}

@misc{Baum2026Anticipatory,
  author = {Baum, Kevin and Kiener, Maximilian and Langer, Markus and Laux, Johann},
  title = {Anticipatory Human Oversight of Agentic {AI}: A Philosophical Account},
  year = {2026},
  eprint = {2609.24242},
  archivePrefix = {arXiv},
  url = {https://arxiv.org/abs/2609.24242}
}

@inproceedings{Malki2026UsableMemory,
  author    = {Malki, Lisa Mekioussa},
  title     = {Towards Usable, Privacy Respecting Long-Term Memory for LLM-based Conversational Agents},
  booktitle = {Extended Abstracts of the 2026 CHI Conference on Human Factors in Computing Systems},
  year      = {2026},
  series    = {CHI EA '26},
  articleno = {909},
  pages     = {1--5},
  publisher = {Association for Computing Machinery},
  doi       = {10.1145/3772363.3799198}
}

@article{Patlan2025ContextManipulation,
  author        = {Patlan, Atharv Singh and Hebbar, Ashwin and Viswanath, Pramod and Mittal, Prateek},
  title         = {Context Manipulation Attacks: Web Agents Are Susceptible to Corrupted Memory},
  journal       = {arXiv preprint arXiv:2506.17318},
  year          = {2025},
  eprint        = {2506.17318},
  archivePrefix = {arXiv},
  primaryClass  = {cs.CR},
  doi           = {10.48550/arXiv.2506.17318}
}

@inproceedings{Sun2025MSA,
  author    = {Sun, Yiheng and Du, Linkang and Su, Zhou and Wang, Yuntao and Liu, Han and Zhao, Quan and Niu, Xiaolin},
  title     = {{MSA}: A Cross-{MCP} Privacy Attack via Memory Exfiltration of Large Language Models},
  booktitle = {Proceedings of the 24th Workshop on Privacy in the Electronic Society},
  year      = {2025},
  pages     = {177--182},
  publisher = {Association for Computing Machinery},
  doi       = {10.1145/3733802.3764057}
}

@inproceedings{Mireshghallah2024KeepSecret,
  author    = {Mireshghallah, Niloofar and Kim, Hyunwoo and Zhou, Xuhui and Tsvetkov, Yulia and Sap, Maarten and Shokri, Reza and Choi, Yejin},
  title     = {Can LLMs Keep a Secret? Testing Privacy Implications of Language Models via Contextual Integrity Theory},
  booktitle = {International Conference on Learning Representations},
  year      = {2024},
  url       = {https://openreview.net/forum?id=gmg7t8b4s0}
}

@techreport{shavit2023practices,
  author      = {Shavit, Yonadav and Agarwal, Sandhini and Brundage, Miles and Adler, Steven and O'Keefe, Cullen and Campbell, Rosie and Lee, Teddy and Mishkin, Pamela and Eloundou, Tyna and Hickey, Alan and Slama, Katarina and Ahmad, Lama and McMillan, Paul and Beutel, Alex and Passos, Alexandre and Robinson, David G.},
  title       = {Practices for Governing Agentic {AI} Systems},
  institution = {OpenAI},
  type        = {White Paper},
  year        = {2023},
  url         = {https://cdn.openai.com/papers/practices-for-governing-agentic-ai-systems.pdf}
}

@inproceedings{turpin2023language,
  author    = {Turpin, Miles and Michael, Julian and Perez, Ethan and Bowman, Samuel R.},
  title     = {Language Models Don't Always Say What They Think: Unfaithful Explanations in Chain-of-Thought Prompting},
  booktitle = {Advances in Neural Information Processing Systems},
  volume    = {36},
  year      = {2023}
}

@article{lanham2023measuring,
  author  = {Lanham, Tamera and Chen, Anna and Radhakrishnan, Ansh and Steiner, Benoit and Denison, Carson and Hernandez, Danny and Li, Dustin and Durmus, Esin and Hubinger, Evan and Kernion, Jackson and others},
  title   = {Measuring Faithfulness in Chain-of-Thought Reasoning},
  journal = {arXiv preprint arXiv:2307.13702},
  year    = {2023}
}

@article{Tran2025PrivacyNorms,
  author  = {Tran, Sarah and Lu, Hongfan and Slaughter, Isaac and Herman, Bernease and Dangol, Aayushi and Fu, Yue and Chen, Lufei and Gebreyohannes, Biniyam and Howe, Bill and Hiniker, Alexis and Weber, Nicholas and Wolfe, Robert},
  title   = {Understanding Privacy Norms Around LLM-Based Chatbots: A Contextual Integrity Perspective},
  journal = {Proceedings of the AAAI/ACM Conference on AI, Ethics, and Society},
  volume  = {8},
  number  = {3},
  pages   = {2522--2534},
  year    = {2025},
  doi     = {10.1609/aies.v8i3.36735}
}

@inproceedings{Li2025PrivaCIBench,
  author    = {Li, Haoran and Hu, Wenbin and Jing, Huihao and Chen, Yulin and Hu, Qi and Han, Sirui and Chu, Tianshu and Hu, Peizhao and Song, Yangqiu},
  title     = {{PrivaCI}-Bench: Evaluating Privacy with Contextual Integrity and Legal Compliance},
  booktitle = {Proceedings of the 63rd Annual Meeting of the Association for Computational Linguistics (Volume 1: Long Papers)},
  month     = jul,
  year      = {2025},
  address   = {Vienna, Austria},
  pages     = {10544--10559},
  publisher = {Association for Computational Linguistics},
  doi       = {10.18653/v1/2025.acl-long.518},
  url       = {https://aclanthology.org/2025.acl-long.518/}
}

@inproceedings{zhou2026operationalizing,
  title={Operationalizing data minimization for privacy-preserving LLM prompting},
  author={Zhou, Jijie and Mireshghallah, Niloofar and Li, Tianshi},
  booktitle={International Conference on Learning Representations},
  volume={2026},
  pages={56919--56943},
  year={2026}
}

@misc{Anthropic2026MemoryEverywhere,
  author       = {{Anthropic}},
  title        = {Claude's Memory Works Everywhere, and You Decide What's in It},
  year         = {2026},
  month        = aug,
  day          = {25},
  howpublished = {Claude by Anthropic},
  url          = {https://claude.com/blog/claudes-memory-works-everywhere-and-you-decide-whats-in-it},
  note         = {Accessed September 9, 2026}
}

@misc{Google2026PersonalIntelligence,
  author       = {{Google}},
  title        = {Connect Your Google Apps to Personalize Your Gemini Experience},
  year         = {2026},
  howpublished = {Gemini Apps Help},
  url          = {https://support.google.com/gemini/answer/16598406},
  note         = {Accessed September 9, 2026}
}

@misc{EDPB2020DataProtectionDefault,
  author       = {{European Data Protection Board}},
  title        = {Guidelines 4/2019 on Article 25 Data Protection by Design and by Default},
  year         = {2020},
  month        = oct,
  day          = {20},
  howpublished = {Final Version},
  url          = {https://www.edpb.europa.eu/documents/guideline/guidelines-42019-on-article-25-data-protection-by-design-and-by-default_en},
  note         = {Accessed September 9, 2026}
}

@article{Herrador2026SpAIware,
  title   = {{SpAIware}: Uncovering a Novel Artificial Intelligence Attack Vector
             through Persistent Memory in {LLM} Applications and Agents},
  author  = {Herrador, Manuel and Rehberger, Johann},
  journal = {Future Generation Computer Systems},
  volume  = {174},
  pages   = {107994},
  year    = {2026}
}

@article{Flanagan1954CIT,
  title   = {The Critical Incident Technique},
  author  = {Flanagan, John C.},
  journal = {Psychological Bulletin},
  volume  = {51}, number = {4}, pages = {327--358}, year = {1954}
}

@inproceedings{Wang2011Regret,
  title     = {``I Regretted the Minute I Pressed Share'': A Qualitative Study of
               Regrets on Facebook},
  author    = {Wang, Yang and Norcie, Gregory and Komanduri, Saranga and
               Acquisti, Alessandro and Leon, Pedro Giovanni and Cranor, Lorrie Faith},
  booktitle = {Proceedings of the Seventh Symposium on Usable Privacy and Security (SOUPS)},
  year      = {2011}
}

@article{Gao2026IsolatedExposed,
  title   = {Isolated but Exposed: Persistence-Based Memory Extraction Attack
             on {LLM} Agents},
  author  = {Gao, Xinyu and Chen, Wenyu and Meng, Xiangtao and Wang, Li and
             Zang, Chuanchao and Wang, Jianing and Li, Zheng and Guo, Shanqing},
  journal = {arXiv preprint arXiv:2607.23444},
  year    = {2026}
}

@misc{zheng2023lmsyschat1m,
  title         = {{LMSYS-Chat-1M}: A Large-Scale Real-World LLM Conversation Dataset},
  author        = {Zheng, Lianmin and Chiang, Wei-Lin and Sheng, Ying and Li, Tianle and Zhuang, Siyuan and Wu, Zhanghao and Zhuang, Yonghao and Li, Zhuohan and Lin, Zi and Xing, Eric P. and Gonzalez, Joseph E. and Stoica, Ion and Zhang, Hao},
  year          = {2023},
  eprint        = {2309.11998},
  archivePrefix = {arXiv},
  primaryClass  = {cs.CL},
  url           = {https://arxiv.org/abs/2309.11998}
}

@article{PalanSchitter2018Prolific,
  author  = {Palan, Stefan and Schitter, Christian},
  title   = {Prolific.ac---A subject pool for online experiments},
  journal = {Journal of Behavioral and Experimental Finance},
  volume  = {17},
  pages   = {22--27},
  year    = {2018},
  doi     = {10.1016/j.jbef.2017.12.004}
}

@article{Finstad2010UMUX,
  author  = {Finstad, Kraig},
  title   = {The Usability Metric for User Experience},
  journal = {Interacting with Computers},
  year    = {2010},
  volume  = {22},
  number  = {5},
  pages   = {323--327},
  doi     = {10.1016/j.intcom.2010.04.004},
  url     = {https://doi.org/10.1016/j.intcom.2010.04.004}
}

@inproceedings{Zhou2025Rescriber,
  author    = {Zhou, Jijie and Xu, Eryue and Wu, Yaoyao and Li, Tianshi},
  title     = {{Rescriber}: Smaller-{LLM}-Powered User-Led Data Minimization for {LLM}-Based Chatbots},
  booktitle = {Proceedings of the 2025 CHI Conference on Human Factors in Computing Systems},
  series    = {CHI '25},
  year      = {2025},
  publisher = {Association for Computing Machinery},
  address   = {New York, NY, USA},
  doi       = {10.1145/3706598.3713701},
  url       = {https://doi.org/10.1145/3706598.3713701}
}

@misc{Zhang2024GhostOfThePast,
  title         = {``Ghost of the past'': Identifying and Resolving Privacy Leakage from LLM's Memory through Proactive User Interaction},
  author        = {Zhang, Shuning and Ye, Lyumanshan and Yi, Xin and Tang, Jingyu and Shui, Bo and Xing, Haobin and Liu, Pengfei and Li, Hewu},
  year          = {2024},
  eprint        = {2410.14931},
  archivePrefix = {arXiv},
  primaryClass  = {cs.HC},
  url           = {https://arxiv.org/abs/2410.14931}
}

@misc{Monteiro2026InferencesAcceptable,
  title         = {When Are LLM Inferences Acceptable? User Reactions and Control Preferences for Inferred Personal Information},
  author        = {Monteiro, Kyzyl and Park, Minjung and Ioffrida, Alexander and Sanna, Angelina and Lee, Hao-Ping and Mireshghallah, Niloofar and Wang, Yang and Das, Sauvik},
  year          = {2026},
  eprint        = {2605.10013},
  archivePrefix = {arXiv},
  url           = {https://arxiv.org/abs/2605.10013}
}

@inproceedings{Yan2025RememberingTrust,
  title     = {Remembering Things Makes Chatbots Sound Smarter, but Less Trustworthy -- A Pilot Study},
  author    = {Yan, Zhicheng and Fischer, Joel E. and Clos, J{\'e}r{\'e}mie},
  booktitle = {Proceedings of the 7th ACM Conference on Conversational User Interfaces},
  series    = {CUI '25},
  year      = {2025},
  publisher = {ACM},
  doi       = {10.1145/3719160.3737617}
}

@article{Malki2025Hoovered,
  title   = {``Hoovered up as a data point'': Exploring Privacy Behaviours, Awareness, and Concerns Among UK Users of LLM-based Conversational Agents},
  author  = {Malki, Lisa Mekioussa and Polamarasetty, Akhil and Hatamian, Majid and Warner, Mark and Costanza, Enrico},
  journal = {Proceedings on Privacy Enhancing Technologies},
  volume  = {2025},
  number  = {4},
  pages   = {838--860},
  year    = {2025},
  doi     = {10.56553/popets-2025-0160}
}

@article{Tamkin2024Clio,
  title   = {Clio: Privacy-Preserving Insights into Real-World {AI} Use},
  author  = {Tamkin, Alex and McCain, Miles and Handa, Kunal and Durmus, Esin and Lovitt, Liane and Rathi, Ankur and Huang, Saffron and Mountfield, Alfred and Hong, Jerry and Ritchie, Stuart and Stern, Michael and Clarke, Brian and Goldberg, Landon and Sumers, Theodore R. and Mueller, Jared and McEachen, William and Mitchell, Wes and Carter, Shan and Clark, Jack and Kaplan, Jared and Ganguli, Deep},
  journal = {arXiv preprint arXiv:2412.13678},
  year    = {2024},
  doi     = {10.48550/arXiv.2412.13678}
}

@article{CostaGomes2025Copilot,
  title   = {It's About Time: The Temporal and Modal Dynamics of {Copilot} Usage},
  author  = {Costa-Gomes, Beatriz and Chen, Sophia and Hsueh, Connie and Morgan, Deborah and Schoenegger, Philipp and Shah, Yash and Way, Sam and Zhu, Yuki and Adeline, Timoth{\'e} and Bhaskar, Michael and Suleyman, Mustafa and Spielman, Seth},
  journal = {arXiv preprint arXiv:2512.11879},
  year    = {2025},
  doi     = {10.48550/arXiv.2512.11879}
}

@inproceedings{Zhao2024WildChat,
  title     = {{WildChat}: 1M {ChatGPT} Interaction Logs in the Wild},
  author    = {Zhao, Wenting and Ren, Xiang and Hessel, Jack and Cardie, Claire and Choi, Yejin and Deng, Yuntian},
  booktitle = {Proceedings of the Twelfth International Conference on Learning Representations (ICLR)},
  year      = {2024},
  url       = {https://openreview.net/forum?id=Bl8u7ZRlbM}
}

@inproceedings{Chen2026AINeverForgets,
  title     = {Relational Gains, Privacy Strains: Exploring Users' Perceptions and Experiences with ChatGPT's Memory Feature},
  author    = {Chen, Cheng and Molina, Maria D. and Liao, Mengqi and Snyder, Eugene Cho},
  booktitle = {Proceedings of the 2026 CHI Conference on Human Factors in Computing Systems},
  series    = {CHI '26},
  year      = {2026},
  publisher = {ACM},
  doi       = {10.1145/3772318.3791635}
}

@inproceedings{Mireshghallah2024TrustNoBot,
  title     = {Trust No Bot: Discovering Personal Disclosures in Human-{LLM} Conversations in the Wild},
  author    = {Mireshghallah, Niloofar and Antoniak, Maria and More, Yash and Choi, Yejin and Farnadi, Golnoosh},
  booktitle = {First Conference on Language Modeling},
  series    = {COLM '24},
  year      = {2024},
  url       = {https://arxiv.org/abs/2407.11438}
}

@inproceedings{Jones2025StorageInterpretation,
  title     = {From Storage to Interpretation: User Perceptions, Practices, and Challenges with Long-term Memory in Agents},
  author    = {Jones, Brennan and Ponochevnyi, Nazar and Stemmler, Kelsey and Su, Emily and Kim, Young-Ho and Kuzminykh, Anastasia},
  booktitle = {Proceedings of the 13th International Conference on Human-Agent Interaction},
  series    = {HAI '25},
  year      = {2025},
  publisher = {ACM},
  doi       = {10.1145/3765766.3765843}
}

@misc{OpenAI2026TemporaryChat,
  author       = {{OpenAI}},
  title        = {Temporary Chat {FAQ}},
  howpublished = {OpenAI Help Center},
  url          = {https://help.openai.com/en/articles/8914046-temporary-chat-faq},
  urldate      = {2026-09-10},
  note         = {Accessed September 10, 2026}
}

@article{acquistiNudgesPrivacySecurity2017,
  title     = {Nudges for {Privacy} and {Security}: {Understanding} and {Assisting} {Users}' {Choices} {Online}},
  author    = {Acquisti, Alessandro and Adjerid, Idris and Balebako, Rebecca and Brandimarte, Laura and Cranor, Lorrie Faith and Komanduri, Saranga and Leon, Pedro Giovanni and Sadeh, Norman and Schaub, Florian and Sleeper, Manya and Wang, Yang and Wilson, Shomir},
  year      = {2017},
  month     = aug,
  journal   = {ACM Computing Surveys},
  volume    = {50},
  number    = {3},
  articleno = {44},
  pages     = {1--41},
  doi       = {10.1145/3054926},
  issn      = {0360-0300},
  url       = {https://doi.org/10.1145/3054926},
  publisher = {Association for Computing Machinery},
  address   = {New York, NY, USA}
}

@article{Wilcoxon1945,
  author  = {Wilcoxon, Frank},
  title   = {Individual Comparisons by Ranking Methods},
  journal = {Biometrics Bulletin},
  year    = {1945},
  volume  = {1},
  number  = {6},
  pages   = {80--83},
  doi     = {10.2307/3001968},
  url     = {https://doi.org/10.2307/3001968}
}

@article{Dunn1961MultipleComparisons,
  author  = {Dunn, Olive Jean},
  title   = {Multiple Comparisons Among Means},
  journal = {Journal of the American Statistical Association},
  year    = {1961},
  volume  = {56},
  number  = {293},
  pages   = {52--64},
  doi     = {10.1080/01621459.1961.10482090},
  url     = {https://doi.org/10.1080/01621459.1961.10482090}
}

@article{Paas1992MentalEffort,
  author  = {Paas, F. G. W. C.},
  title   = {Training Strategies for Attaining Transfer of Problem-Solving Skill in Statistics: A Cognitive-Load Approach},
  journal = {Journal of Educational Psychology},
  year    = {1992},
  volume  = {84},
  number  = {4},
  pages   = {429--434},
  doi     = {10.1037/0022-0663.84.4.429},
  url     = {https://doi.org/10.1037/0022-0663.84.4.429}
}

@inproceedings{Koerber2019Trust,
  author    = {K{\"o}rber, Moritz},
  title     = {Theoretical Considerations and Development of a Questionnaire to Measure Trust in Automation},
  editor    = {Bagnara, Sebastiano and Tartaglia, Riccardo and Albolino, Sara and Alexander, Thomas and Fujita, Yushi},
  booktitle = {Proceedings of the 20th Congress of the International Ergonomics Association ({IEA} 2018)},
  series    = {Advances in Intelligent Systems and Computing},
  volume    = {823},
  year      = {2019},
  pages     = {13--30},
  publisher = {Springer},
  address   = {Cham},
  doi       = {10.1007/978-3-319-96074-6_2},
  url       = {https://doi.org/10.1007/978-3-319-96074-6_2}
}

@misc{Anthropic2026Incognito,
  author       = {{Anthropic}},
  title        = {Use Incognito Chats},
  year         = {2026},
  howpublished = {Claude Help Center},
  url          = {https://support.claude.com/en/articles/12260368-use-incognito-chats},
  note         = {Accessed September 6, 2026}
}

@inproceedings{Chhikara2025Mem0,
  author    = {Chhikara, Prateek and Khant, Dev and Aryan, Saket and Singh, Taranjeet and Yadav, Deshraj},
  title     = {Mem0: Building Production-Ready AI Agents with Scalable Long-Term Memory},
  booktitle = {ECAI 2025},
  pages     = {2993--3000},
  year      = {2025},
  doi       = {10.3233/FAIA251160}
}

@inproceedings{Xu2025AMem,
  author    = {Xu, Wujiang and Liang, Zujie and Mei, Kai and Gao, Hang and Tan, Juntao and Zhang, Yongfeng},
  title     = {A-Mem: Agentic Memory for LLM Agents},
  booktitle = {Advances in Neural Information Processing Systems},
  volume    = {38},
  year      = {2025},
  doi       = {10.52202/085713-0593}
}

@inproceedings{zhang2024fair,
author = {Zhang, Zhiping and Jia, Michelle and Lee, Hao-Ping (Hank) and Yao, Bingsheng and Das, Sauvik and Lerner, Ada and Wang, Dakuo and Li, Tianshi},
title = {“It's a Fair Game”, or Is It? Examining How Users Navigate Disclosure Risks and Benefits When Using LLM-Based Conversational Agents},
year = {2024},
isbn = {9798400703300},
publisher = {Association for Computing Machinery},
address = {New York, NY, USA},
url = {https://doi.org/10.1145/3613904.3642385},
doi = {10.1145/3613904.3642385},
booktitle = {Proceedings of the 2024 CHI Conference on Human Factors in Computing Systems},
articleno = {156},
numpages = {26},
location = {Honolulu, HI, USA},
series = {CHI '24}
}

@inproceedings{habibTogglesDollarSigns2021,
  title = {Toggles, {{Dollar Signs}}, and {{Triangles}}: {{How}} to ({{In}}){{Effectively Convey Privacy Choices}} with {{Icons}} and {{Link Texts}}},
  shorttitle = {Toggles, {{Dollar Signs}}, and {{Triangles}}},
  booktitle = {Proceedings of the 2021 {{CHI Conference}} on {{Human Factors}} in {{Computing Systems}}},
  author = {Habib, Hana and Zou, Yixin and Yao, Yaxing and Acquisti, Alessandro and Cranor, Lorrie and Reidenberg, Joel and Sadeh, Norman and Schaub, Florian},
  year = 2021,
  month = may,
  pages = {1--25},
  publisher = {ACM},
  address = {Yokohama Japan},
  doi = {10.1145/3411764.3445387},
  urldate = {2026-09-10},
  isbn = {978-1-4503-8096-6},
  langid = {english}
}

@inproceedings{Nielsen1994,
  author    = {Nielsen, Jakob},
  title     = {Enhancing the Explanatory Power of Usability Heuristics},
  booktitle = {Proceedings of the SIGCHI Conference on Human Factors in Computing Systems},
  year      = {1994},
  pages     = {152--158},
  publisher = {Association for Computing Machinery},
  address   = {New York, NY, USA},
  doi       = {10.1145/191666.191729}
}

@inproceedings{Kulesza2013TooMuch,
  author    = {Kulesza, Todd and Stumpf, Simone and Burnett, Margaret and Yang, Sherry and Kwan, Irwin and Wong, Weng-Keen},
  title     = {Too Much, Too Little, or Just Right? Ways Explanations Impact End Users' Mental Models},
  booktitle = {2013 IEEE Symposium on Visual Languages and Human-Centric Computing (VL/HCC)},
  pages     = {3--10},
  year      = {2013},
  organization = {IEEE},
  doi       = {10.1109/VLHCC.2013.6645235}
}

@inproceedings{Kizilcec2016,
  author    = {Kizilcec, Ren{\'e} F.},
  title     = {How Much Information?: Effects of Transparency on Trust in an Algorithmic Interface},
  booktitle = {Proceedings of the 2016 CHI Conference on Human Factors in Computing Systems},
  pages     = {2390--2395},
  year      = {2016},
  publisher = {ACM},
  address   = {San Jose, CA, USA}
}

@inproceedings{Muller2022Noticeability,
  author    = {M{\"u}ller, Philipp and Staal, Sander and B{\^a}ce, Mihai and Bulling, Andreas},
  title     = {Designing for Noticeability: Understanding the Impact of Visual Importance on Desktop Notifications},
  booktitle = {Proceedings of the 2022 CHI Conference on Human Factors in Computing Systems},
  year      = {2022},
  pages     = {1--13},
  publisher = {Association for Computing Machinery}
}

@article{Song2026DontLook,
  author  = {Song, Ji Eun and Lee, Eunchae and Im, Juhee and Jang, Hyunsoo and Kim, Eunji and Lee, Joongseek},
  title   = {"Don't Look, But I Know You Do": Norms and Observer Effects in Shared LLM Accounts},
  journal = {arXiv preprint arXiv:2603.22822},
  year    = {2026}
}

@misc{Phang2025Affective,
  author = {Phang, Jason and Lampe, Michael and Ahmad, Lama and Agarwal, Sandhini and Fang, Cathy Mengying and Liu, Auren R. and Danry, Valdemar and Lee, Eunhae and Chan, Samantha W. T. and Pataranutaporn, Pat and Maes, Pattie},
  title = {Investigating Affective Use and Emotional Well-being in ChatGPT},
  year = {2025}
}
